%% file: su3monopoletda.tex
\documentclass[prl,aps,
longbibliography,
preprintnumbers,
twocolumn,
nofootinbib,
floatfix,
superscriptaddress,amsmath,amssymb]{revtex4-2}

\usepackage{graphicx}
\usepackage{xcolor}
\usepackage{dcolumn}
\usepackage{bm}
\usepackage[T1]{fontenc} 
\usepackage{epsfig}
\usepackage{rotating}
\usepackage{amssymb}
\usepackage{dsfont}
\usepackage{psfrag}
\usepackage{amsmath,euscript,array,mathrsfs}
\usepackage{bbold}
\usepackage{epsf}
\usepackage{float}
\floatstyle{plaintop}
\usepackage[utf8]{inputenc}
\usepackage{placeins}
\usepackage{orcidlink}
\usepackage{soul}
\usepackage{xr} 
\usepackage[caption=false]{subfig}

\newcommand{\beq}{\begin{equation}}
\newcommand{\eeq}{\end{equation}}
\newcommand{\beqs}{\begin{eqnarray}}
\newcommand{\eeqs}{\end{eqnarray}}

\newcommand{\Tr}{{\rm Tr}}

\newcommand{\orcidauthorCREAN}{0009-0007-8620-6243}
\newcommand{\orcidauthorGIANSIRACUSA}{0000-0003-4252-0058}
\newcommand{\orcidauthorLUCINI}{0000-0001-8974-8266}

\begin{document}

\title{Simplicity of confinement in SU(3) Yang-Mills theory}

\author{Xavier Crean\,\orcidlink{\orcidauthorCREAN}}
\email{2237451@swansea.ac.uk}
\affiliation{Department of Mathematics, Faculty  of Science and Engineering,
Swansea University, Fabian Way, Swansea, SA1 8EN, UK}

\author{Jeffrey Giansiracusa\,\orcidlink{\orcidauthorGIANSIRACUSA}}
\email{jeffrey.giansiracusa@durham.ac.uk}
\affiliation{Department of Mathematical Sciences, Durham University, Upper Mountjoy Campus, Durham, DH1
3LE, UK}

\author{Biagio Lucini\,\orcidlink{\orcidauthorLUCINI}}
\email{b.lucini@qmul.ac.uk}
\affiliation{School of Mathematical Sciences, Queen Mary University of London, Mile End Road, London, E1 4NS, UK}

\date{\today}%
\begin{abstract}
We introduce a novel observable associated to Abelian monopole currents defined in the
Maximal Abelian Projection of SU(3) Yang-Mills theory that captures the
topology of the current loop. This observable, referred to as
the {\em simplicity}, is defined as the ratio of the zeroth over the 
first Betti number of the current graph for a given field
configuration. A numerical study of the  expectation value of the
simplicity performed in the framework of Lattice Gauge Theories enables 
us to determine the deconfinement temperature to a higher degree of
accuracy than that reached by conventional methods at a comparable
computational effort. Our results suggest that Abelian
current loops are strongly correlated with the degrees of freedoms of
the theory that determine confinement. Our investigation opens new
perspectives for the definition of an order parameter for
deconfinement in Quantum Chromodynamics able to expose the potentially
rich phase structure of the theory.  
\end{abstract}

\maketitle

\noindent
{\em Introduction --- }
The problem of colour confinement in the strong interactions, i.e.,
the absence of fractionally charged particles in asymptotic states, is
broadly acknowledged as one of the most urgent gaps to fill in order to gain a
full understanding of Quantum Chromodynamics (QCD) and, more in
general, of the dynamics of non-Abelian gauge theories. It is known
that at high temperature confinement is lost. The widely accepted
scenario (for a recent review, see, e.g., Ref.~\cite{Aarts:2023vsf}) is that
above a critical temperature the system behaves as a plasma of its
elementary constituents, quarks and gluons. This picture has recently
been challenged, and the suggestion that an intermediate regime
separates the confinement regime from the deconfined
quark-gluon-plasma one (see, e.g.,
Refs.~\cite{Glozman:2016swy,Alexandru:2019gdm,Cardinali:2021mfh,Hanada:2023krw,Fujimoto:2025sxx,Mickley:2025qgk})
is gaining increasing consensus. In the light of this development, 
a full characterisation of confinement appears to be even more crucial. 
To study QCD, a first-principles approach that has been used, often in conjunction
with analytic guidance, is Lattice Gauge Theory,
of which the demonstration of confinement through the area law of Wilson loops has been the
very first application \cite{Wilson:1974sk}.

It is an old idea that
confinement can be understood in terms of the dynamics of topological
excitations carrying a non-trivial magnetic charge
\cite{Mandelstam:1974pi,tHooft:1975yol}. This has led to the proposal
of defining Abelian magnetic monopoles through a partial gauge fixing
procedure known as {\em Abelian projection}
\cite{tHooft:1981bkw}. A number of studies were
performed which explored these ideas from various angles
\cite{Kronfeld:1987vd,Kronfeld:1987ri,Ivanenko:1991wt}, from
dominance of degrees of freedom that are Abelian in certain
projections~\cite{Ezawa:1982bf} to the
construction of disorder parameters
\cite{DelDebbio:1995yf,DiGiacomo:1999yas,DiGiacomo:1999fb}\footnote{Pathologies identified in the
original disorder parameter construction~\cite{Greensite:2008ss} led
to the improved proposal~\cite{Bonati:2011jv}.} and
effective monopole potentials
\cite{Chernodub:1996ps}. Studies of the
gauge-independence of monopole condensation \cite{Carmona:2001ja} and
investigations of thermal monopole properties
\cite{Chernodub:2006gu,DAlessandro:2007lae} provided important
insights into the connection between monopole dynamics and the
confinement/deconfinement transition, with an approach based on an
effective model for the condensation of monopoles at the critical
temperature~\cite{Cristoforetti:2009tx} enabling the computation of the latter in SU(2)
Yang-Mills~\cite{DAlessandro:2010jdd}. More recent
investigations~\cite{Nguyen:2024ikq,Giansiracusa:2025wqn} reaffirmed
the centrality of monopoles in colour confinement. 

Despite this noticeable progress, we still lack a robust (dis)order
parameter for confinement in Yang-Mills theory based on features of
these topological excitations that is free from
lattice artefacts, enables determining with sufficient accuracy
quantities that characterise the dynamics of the transition, and can
be readily adopted in the presence of dynamical fermions.
A key difficulty is the topological nature of monopole excitations,
which is harder to expose in the lattice discretisation. Recently,
Topological Data Analysis (TDA) has emerged in computational topology
as a robust methodology to characterise topological properties of
discrete sets of points. For this reason, applications of TDA to field
theory and statistical mechanics are raising at a fast pace (see, e.g.,
\cite{topology-hypothesis-with-PH,Tran:2021wyg,Kashiwa:2021ctc,Olsthoorn:2020xzs,Cole:2020hjx,Sale:2021xsq,Sehayek:2022lxf,Sale:2022qfn,Spitz:2022tul,Spitz:2024bqh}).   
Building on the intuition developed from previous studies and
borrowing rigorous tools of TDA along the
path laid in Refs.~\cite{Crean:2024nro,Crean:2025gne}, in this
letter we are going to propose and test in SU(3) Yang-Mills an
observable derived from magnetic monopole currents that quantitatively
characterises the deconfinement phase transition. We will demonstrate
numerically that this observable, whose definition in the presence of
fermions remains identical, enables us to accurately determine the
critical value of the coupling and the order of the phase transition
in the pure gauge system, henceforth providing evidence of the
relevance of Abelian monopoles for deconfinement in Yang-Mills.  

{\em SU(3) Yang-Mills on the Lattice --- }
We consider the SU(3) lattice gauge theory described by the Wilson action,
\begin{equation}\label{eq:wilson-action}
    S = \beta \sum_{i} \sum_{\mu < \nu} \left[ 1 - \frac{1}{3} \Re
      \mathrm e \Tr U_{\mu \nu} (i) \right],
\end{equation}
with $\beta \equiv 6 / g^{2}$ and $g$ the coupling of the theory. This
action becomes the Yang-Mills action in the continuum limit. The quantity
\begin{equation}\label{eq:plaquette-wilson-loop}
    U_{\mu\nu} (i) \equiv U_{\mu}(i)U_{\nu}(i + \hat{\mu})U_{\mu}^{\dag}(i+\hat{\nu})U_{\nu}^{\dag}(i)  
  \end{equation}
 is the plaquette variable, with $U_{\mu}(i) \in$ SU(3) the link
 variable, defined on the link $(i; \hat{\mu})$ stemming from the point $i$
 and ending at the point $i + \hat{\mu}$ (with $\hat{\mu}$ the unit
 vector in the positive direction $\mu$) of a four-dimensional
 Euclidean lattice of dimension $N_t \times N_s^3$.  Periodic boundary
 conditions are imposed in all directions. The temperature $T$ of the
 system is given by 
\begin{equation}\label{eq:temperature-temporal-direction}
    T = \frac{1}{a(\beta) N_t} \ , 
\end{equation}
with $a(\beta)$ the lattice spacing, which is a monotonically decreasing function of $\beta$.
The path integral is given by 
  \begin{equation}\label{eq:partition-function}
    Z = \int \left( \prod_{i, \mu} d U_{\mu}(i) \right) \ \exp \{ -S \},
\end{equation}
with $d U_{\mu}(i)$ representing the Haar measure.

At thermal equilibrium, the expectation value of a given observable $O$ is
computed using   
\begin{equation}\label{eq:thermal-expectation-value}
    \langle O \rangle = \frac{1}{Z} \int  \left( \prod_{i,\mu}
      d U_{\mu}(i) \right) \,  O \exp \{ -S \}.
\end{equation}
By fixing $N_{t}$ and taking the infinite volume limit $N_{s} \to
\infty$, we can study the system's behaviour in the thermodynamic
limit. The thermodynamic observables thus extracted can then be
extrapolated to the continuum limit by taking progressively finer
discretisations in $N_t$. 

{\em Magnetic monopoles in the Maximal Abelian Gauge --- }
The identification of Abelian magnetic monopole currents in $\mathrm{SU}(N)$
gauge theories is performed following a gauge-fixing procedure known
as an {\em Abelian projection}. To define an Abelian projection, an
operator $F$ transforming in the adjoint representation is chosen, and
a partial gauge fixing is performed. The latter consists in
diagonalising $F$ at each point and ordering its eigenvalues
$\lambda_1, \dots \lambda_N$ in non-decreasing order. This partial
gauge fixing leaves a residual $\mathrm{U}(1)^{N -1}$ symmetry. The
process exposes $N-1$ species of magnetic monopoles, each corresponding
to one of the $\mathrm{U}(1)$ residual gauge factors. Magnetic
monopoles of the type $j$ occur where $\lambda_j = \lambda_{j+ 1}$. In
$\mathrm{SU}(3)$ Yang-Mills, we then have two types of monopoles. 

While the gauge-dependence of the identification of monopole currents
might obscure the gauge-invariant picture, physical properties of
magnetic monopoles in some gauge can be a useful starting point to
better understand properties of more complex underlying topological
structures that the gauge-dependent objects exemplify. In this
respect, a particularly useful gauge is the Maximal Abelian
Gauge ({\em MAG}), since gauge-independent excitations maximally overlap with
Abelian monopoles defined in this gauge~\cite{Bonati:2010tz}. 
Following the prescription of Ref.~\cite{Bonati:2013bga}, for the $\mathrm{SU}(3)$ lattice
Yang-Mills theory, we define the adjoint operator $\tilde X(i)$ as 
\begin{widetext}
\begin{equation}
\tilde X(i) = \sum_\mu \left[ U_\mu(i) \tilde\lambda U^\dagger_\mu(i)
+U^\dagger_\mu(i-\hat{\mu}) \tilde\lambda U_\mu(i-\hat{\mu})\right] \ , 
\qquad \tilde \lambda = {\rm diag} (1,0,-1) \ .  
\end{equation}
The MAG is defined as the gauge in which $\tilde X(i)$ is diagonal. This gauge choice is equivalent to requiring that the operator 
\begin{eqnarray}
\tilde F_{\rm MAG} (U,g ) = \sum_{\mu,i} 
\mbox{tr} \left( g(i) U_\mu(i) g^{\dagger}  (i + \hat{\mu})  \tilde\lambda g (i + \hat{\mu}) U^{\dagger}_\mu(i) g^\dagger(i) \, \tilde\lambda \right) 
\end{eqnarray}
\end{widetext}
is maximised over $g$, i.e., the gauge fixing transformation $\{ g \}$ can be derived from the condition
\begin{eqnarray}
\{ \tilde g  \}  = \mathop{\rm argmax}_{\{ g \} } \tilde F _{\rm MAG}
  ( U  , g ) \ .
\end{eqnarray}
In the MAG, the diagonal elements of the link matrices $\tilde U _{i i}$ read
\begin{eqnarray}
\tilde U _{i i} = r_i e^{i \varphi_i} \ , \qquad \sum_i \varphi_i = 2
  \pi n + \delta \varphi \ .
\end{eqnarray}
In general, $\delta \varphi \ne 0$, as a consequence of the fact that
even after gauge fixing the links are not fully diagonal. 
Angle variables $\phi_i$ are then defined through the redistribution of the excess phase as
\begin{eqnarray}
\phi_i = \varphi_i - \delta  \varphi\frac{ \left| \tilde U_{ii} \right |^{-1}}{\sum_j \left| \tilde U_{jj}\right| ^{-1}} \ .
\end{eqnarray}
The two lattice Abelian fields in the residual gauge are $\theta_1 =
\phi_1$ and $\theta_2 = - \phi_3$. The corresponding species of
monopoles are defined following the prescription of
Ref.~\cite{DeGrand:1980eq}. The truncation of the theory to the degrees of
freedom represented by these two angles provides a definition of {\em
  Abelian projection} using the MAG. 

{\em Topological simplicity --- }
To investigate the topology of monopole currents, it is convenient to
define the dual lattice $\Lambda^{\ast}$, 
obtained by shifting each point by half a lattice spacing in all
positive directions. Each geometric element of the original lattice $\Lambda$ of dimension
$d$ is pierced at its centre by an element of $\Lambda^{\ast}$ of
dimension $4 -d$. These two elements are dual to each
other. Being defined on cubes of $\Lambda$, on $\Lambda^{\ast}$
magnetic charges are link variables and are more appropriately
referred to as currents. For each lattice configuration, each dual
link $(j; \hat{\mu})$ carries a current $m_{\mu}(j)  = 0, \pm 1, \pm 2$. 
A non-trivial current $m$ on the dual link
$(j; \hat{\mu} )$ can be seen as a charge $m$ moving from site $j$ to
site $j + \hat{\mu}$. Akin to Kirchoff's law, the total charge
entering a dual site $j$ is equal to the total charge exiting from
that site. For the same reason, currents cannot end at sites but
must form closed loops. The set of links $G \equiv \{(j; \hat{\mu}), \, 
m_\mu(j) \ne 0 \}$ is the current graph associated to the
configuration on which the $m_{\mu}(j)$ have been computed. This
object is interpreted as a graph, with vertices coinciding with the
end points of links in $G$ and edges coinciding with the links in $G$. The number of connected components of $G$ defines the Betti
number $b_0$, while the total number of loops defines the Betti number
$b_1$.  In~\cite{Crean:2025gne}, we have shown that both quantities
\begin{eqnarray}
\rho_0 = \frac{\langle b_0 \rangle}{N_s^3} \qquad \mbox{and} \qquad
  \rho_1= \frac{\langle b_1 \rangle}{N_s^3}
\end{eqnarray}
are sensitive to the phase transition, by providing numerical
evidence that their susceptibilities, $\chi_0$ and $\chi_1$, display a
peak at a value $\beta_c(N_s, N_t)$ of $\beta$ that for fixed $N_t$ scales as
\begin{eqnarray}
\label{eq:scalingfirstorder}
  \beta_c(N_s,N_t) = \beta_c(N_t) + a/N_s^3 \ , 
\end{eqnarray}
where $\beta_c(N_t)$ is the critical value of $\beta$ at fixed $N_t$. 
While these results show a remarkable connection between topological
properties of current graphs and the deconfinement phase transition,
suggesting that expectation values of the Betti numbers $b_0$ and
$b_1$ are key to understanding confinement, they fail short of
providing an order parameter for the deconfinement 
phase transition. 

In this work, we use the information above and expectations
developed from previous calculations to define a quantity that behaves
as an order parameter. This observable is the {\em topological simplicity},
or just {\em simplicity}, which is defined as
\begin{eqnarray}
  \lambda = \left\langle \frac{b_0}{b_1} \right\rangle \ .
\end{eqnarray}
The ratio that defines this observable, $\frac{b_0}{b_1} $, is the
reciprocal of the number of loops per connected component, which is sometimes
called the \emph{complexity} and is at least 1.  While the simplicity and complexity are
undefined for an empty network, a typical configuration at non-infinite $\beta$
will have a non-empty monopole network. In the confined phase at low temperature
(corresponding at fixed $N_t$ to low $\beta$), one typically sees a percolating
current network with one component and many loops, and so the simplicity
approaches 0. In the deconfined phase at high temperature (corresponding at
fixed $N_t$ to high $\beta$), large fluctuations are suppressed by the Boltzmann
weight and dominant configurations have current networks consisting of a sparse
gas of loops, so the simplicity is approximately 1. In the critical
region, due to the breaking of the percolating loop, the simplicity
raises from 0 to 1.

\begin{table}[]
    \centering
    \renewcommand{\arraystretch}{1.5}
    \begin{tabular}{|c|c|c|c|c|c|c|}
        \hline
        $N_t$ & $N_s$ & $\beta_{\mathrm{min}}$ &
                $\beta_{\mathrm{max}}$&  $N_{\beta}$ & $N_{\mathrm{meas}}$\\
        \hline
        \hline
        $4$ & $16$, $20$, $24$, $28$, $32$  & $5.6600$ & $5.7200$ & 15 & 600\\
        $6$ & $24$, $30$, $36$, $42$, $48$  & $5.8100$ & $5.9300$ & 12 & 600\\
        $8$ & $32$, $40$, $48$, $56$, $64$  & $6.0350$ & $6.0800$ & 12 & 400\\
        \hline
    \end{tabular}
    \caption{This table specifies the various lattice sizes, $N_t$ and
      $N_s$, that we use in this study. For $N_t = 4$, we draw a
      sample of $N=600$ configurations, for a range of $15$ $\beta$
      values. For $N_t = 6$, we draw a sample of $N=600$
      configurations, for a range of $12$ $\beta$ values. For $N_t =
      8$, we draw a sample of $N=400$ configurations, for a range of
      $12$ $\beta$ values.} 
    \label{tab:lattice-sizes}
\end{table}

\begin{figure}
    \centering
    \includegraphics[width=0.9\columnwidth]{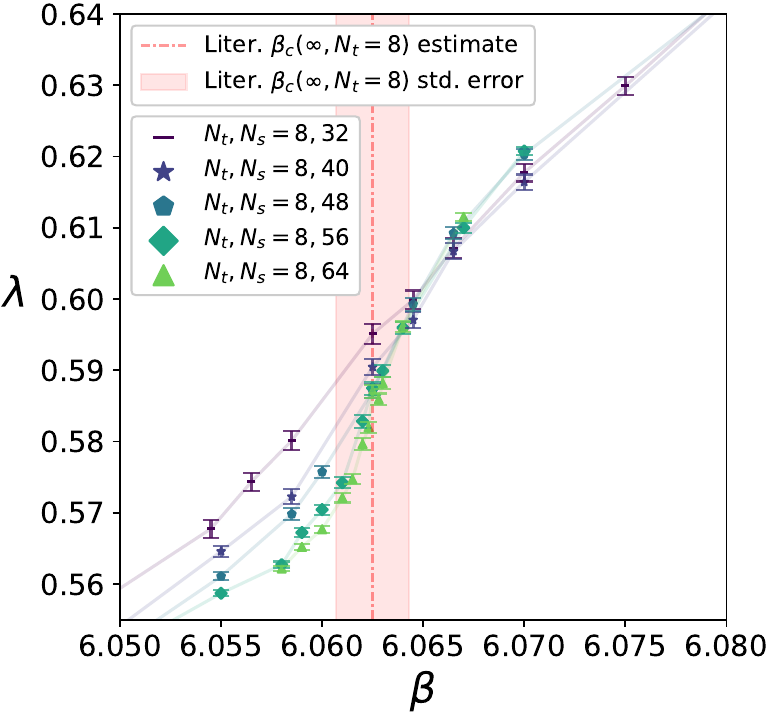}
    \includegraphics[width=0.9\columnwidth]{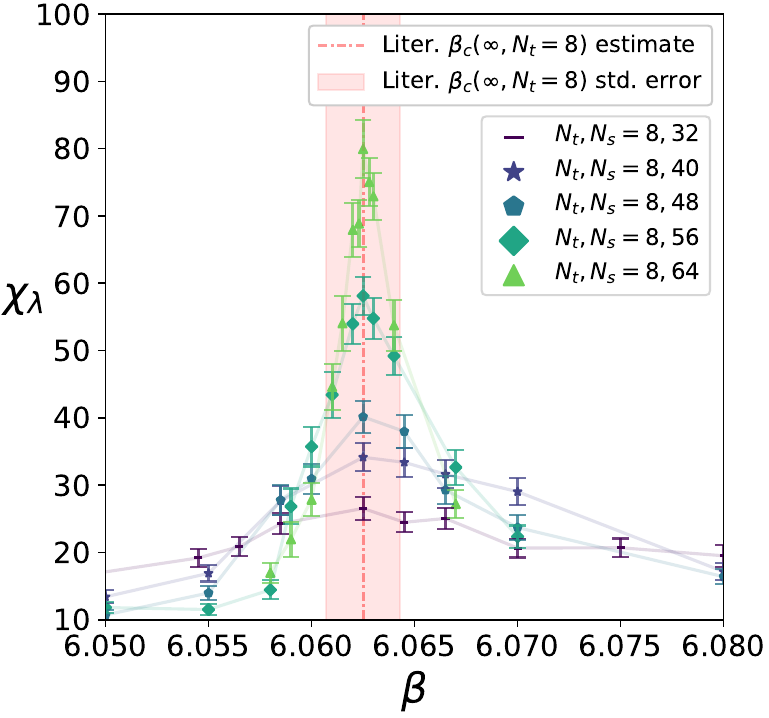}
    \caption{For $N_t = 8$, $\lambda$ and
      $\chi_\lambda$ as functions of $\beta$ zoomed into the critical
      region with translucent lines to guide the eye. The vertical line and band show respectively the central value and the statistical error for the extrapolated $\beta_{c}$ determined in Ref.~\cite{Lucini:2003zr}.}
    \label{fig:Nt=8_scatter}
\end{figure}

\begin{figure}
    \centering
    \includegraphics[width=0.95\columnwidth]{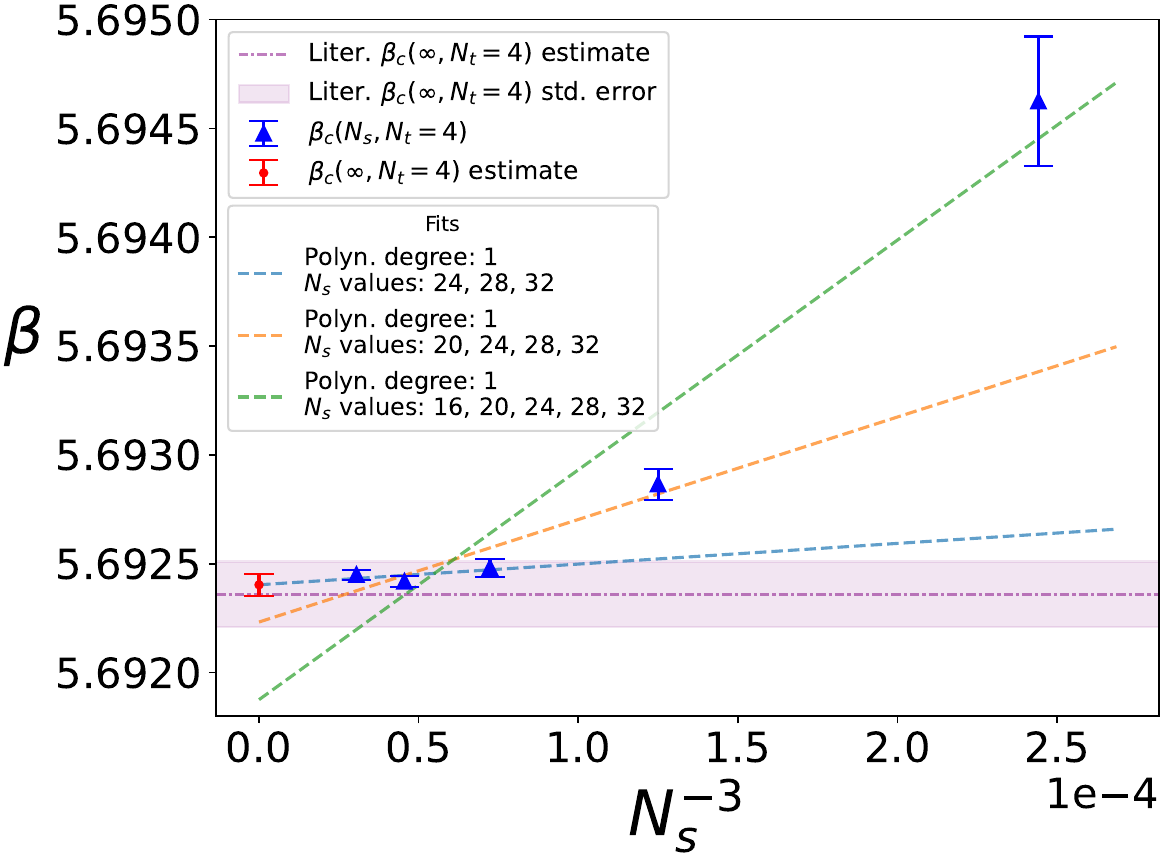}
    \caption{ Position of the peak of $\chi_\lambda$ for lattices with
      $N_t = 4$ and $N_s = 16,20,24,28,32$ (blue triangles). Dashed
      lines represent the polynomial regression fits used for the
      infinite volume extrapolation of these peak values, and the red circle is
      the resulting estimate of $\beta_{c}$ in the thermodynamic limit
      (see the supplemental material for additional details). The
      horizontal line and band show respectively the central value and
      the statistical error for the extrapolated 
      $\beta_{c}$ determined in Ref.~\cite{Lucini:2003zr}.} 
    \label{fig:fits_ratio_Nt4}
\end{figure}
\begin{figure}
    \centering
    \includegraphics[width=0.95\columnwidth]{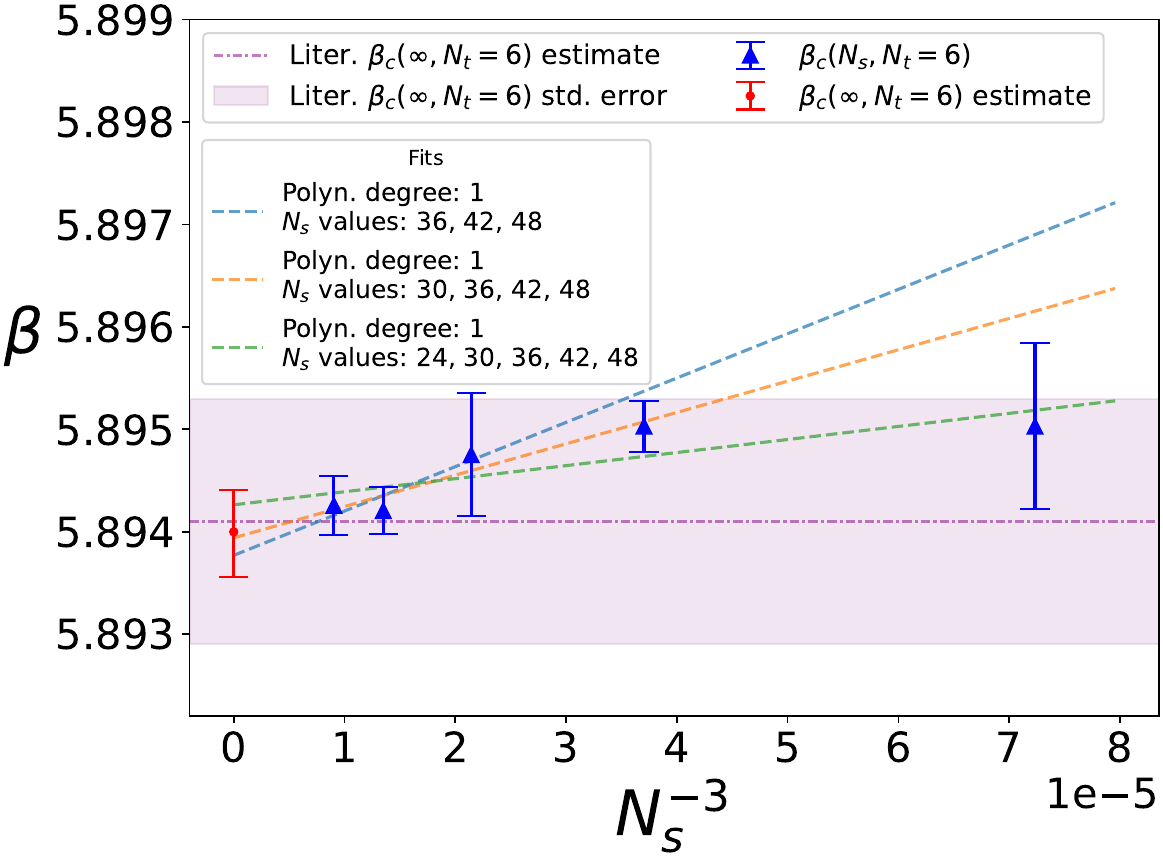}
    \caption{Position of the peak of $\chi_\lambda$ for lattices with $N_t = 6$
      and $N_s = 24,30,36,42,48$ (blue triangles). Dashed
      lines represent the polynomial regression fits used for the
      infinite volume extrapolation of these peak values, and the red circle is
      the resulting estimate of $\beta_{c}$ in the thermodynamic limit
      (see the supplemental material for additional details). The
      horizontal line and band show respectively the central value and
      the statistical error for the extrapolated 
      $\beta_{c}$ determined in Ref.~\cite{Lucini:2003zr}.}
    \label{fig:fits_ratio_Nt6}
\end{figure}
\begin{figure}
    \centering
    \includegraphics[width=1.0\columnwidth]{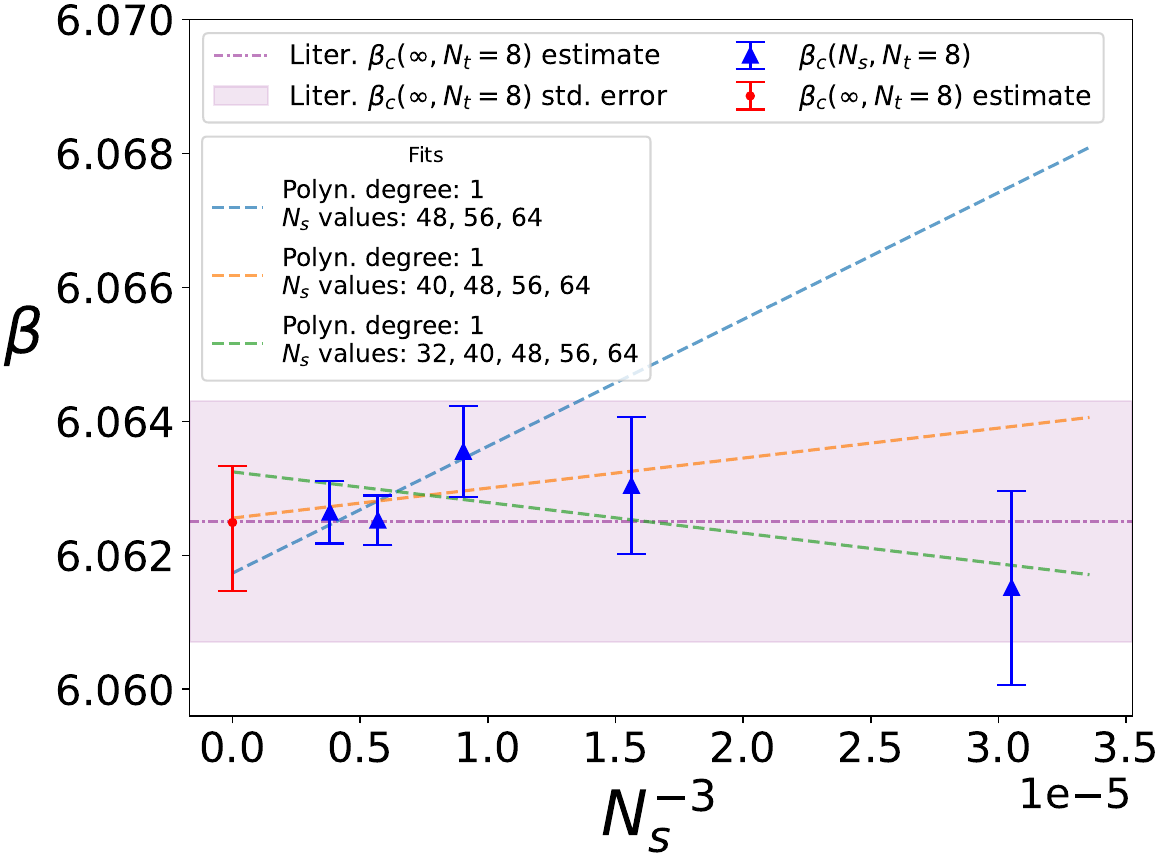}
    \caption{Position of the peak of $\chi_\lambda$ for lattices with
      $N_t = 8$ and $N_s = 32,40,48,56,64$ (blue triangles). Dashed
      lines represent the polynomial regression fits used for the
      infinite volume extrapolation of these peak values, and the red circle is
      the resulting estimate of $\beta_{c}$ in the thermodynamic limit
      (see the supplemental material for additional details). The
      horizontal line and band show respectively the central value and
      the statistical error for the extrapolated $\beta_{c}$
      determined in Ref.~\cite{Lucini:2003zr}.}
    \label{fig:fits_ratio_Nt8}
\end{figure}

However, this qualitative behaviour is not sufficient to evidence the
coupling of $\lambda$ with the degrees of freedom that are relevant for
the phase transition. In order to test whether $\lambda$ is an
observable that is sensitive to the deconfinement phase transition, we
study the behaviour of its susceptibility
\begin{eqnarray}
\chi_\lambda = V \left( \left\langle \left( \frac{b_0}{b_1} \right)^2 \right\rangle
  - \left\langle \frac{b_0}{b_1} \right\rangle ^2 \right) 
\end{eqnarray} 
in a range of $\beta$ values in the critical region, for a range of spatial
extensions $N_s$ and temporal extensions $N_t = 4, 6, 8$, as specified in
Tab.~\ref{tab:lattice-sizes}. Larger values of $N_t$
correspond to finer discretisations of the direction
associated with the temperature (see
Eq.~\eqref{eq:temperature-temporal-direction}). Varying $N_s$ will
instead enable us to prove sensitivity to the expected first-order
deconfinement phase transition, which will be shown by the detection
of the scaling behaviour in Eq.~\eqref{eq:scalingfirstorder}. 

We report in Fig.~\ref{fig:Nt=8_scatter} the behaviour of $\lambda$
and of $\chi_{\lambda}$ together with a determination for
$\beta_c$ for $N_t = 8$ with conventional methods taken from
Ref.~\cite{Lucini:2003zr}. $\lambda$ shows a monotonically
increasing behaviour, with slope that gets steeper near $\beta_c$ and
becomes more pronounced as the volume increases. As a consequence, the
susceptibility $\chi_{\lambda}$ has a peak in the critical region with
height growing with the
volume. Figs.~\ref{fig:fits_ratio_Nt4},~\ref{fig:fits_ratio_Nt6}~and~\ref{fig:fits_ratio_Nt8}
show the scaling of the position of the peak of
$\chi_{\lambda}$ with the spatial volume, respectively at $N_t =
4,6,8$. The data have been fitted with
Eq.~\eqref{eq:scalingfirstorder} and with a higher-order 
correction in $1/N_s^3$ for various fitting ranges, and the results
have been combined with the Akaike information principle in the
implementation of Ref.~\cite{Jay:2020jkz} to determine a weighted
extrapolation and the corresponding errors (both also reported in the
figures). 

{\em Discussion and outlook ---} The smaller error bar of the
extrapolated value with respect to the reference value, which has been obtained by averaging over around an order of
magnitude more configurations, provides evidence of the relevance of
the topological properties captured by the simplicity for the dynamics
of colour confinement. We remark that the simplicity can be adopted
also in the presence of
dynamical fermions. If a novel intermediate-temperature behaviour exists in
QCD, based on our studies of spin systems, where
observables constructed with similar methodology as the simplicity
have proved to be sensitive to the order-disorder phase transition, we
expect the susceptibility of the simplicity to show a non-monotonic
structure at the higher-temperature change of regime. 

\acknowledgments

{\em Acknowledgements ---} We thank M. D'Elia, A. Gonzalez-Arroyo and
T. Sulejmanpasic for discussions. XC was supported by the Additional
Funding Programme for Mathematical Sciences, delivered by EPSRC
(EP/V521917/1) and the Heilbronn Institute for Mathematical
Research. JG was supported by EPSRC grant EP/R018472/1 through the
Centre for TDA and the Erlangen Hub for AI through EPSRC grant
EP/Y028872/1. The work of BL was partly supported by the EPSRC
ExCALIBUR ExaTEPP project EP/X017168/1 and by the STFC Consolidated
Grants No. ST/T000813/1 and ST/X000648/1. 

Numerical simulations have been performed on the Swansea SUNBIRD cluster (part of the Supercomputing Wales project) and AccelerateAI A100 GPU system. The Swansea SUNBIRD system and AccelerateAI are part funded by the European Regional Development Fund (ERDF) via Welsh Government. 

This work used the DiRAC Data Intensive service (CSD3) at the University of Cambridge, managed by the University of Cambridge University Information Services on behalf of the STFC DiRAC HPC Facility (www.dirac.ac.uk). The DiRAC component of CSD3 at Cambridge was funded by BEIS, UKRI and STFC capital funding and STFC operations grants. DiRAC is part of the UKRI Digital Research Infrastructure.

{\bf Open Access Statement} --- For the purpose of open access, the authors have applied a Creative Commons Attribution (CC BY) licence  to any Author Accepted Manuscript version arising.

{\bf Research Data Access Statement ---} The data and analysis code for
this manuscript can be downloaded from  Ref.~\cite{ZENODO1}. The
Monte Carlo code can be found from Ref.~\cite{ZENODO2}. 

\bibliographystyle{apsrev4-2}
\bibliography{refs_new}

\clearpage
\onecolumngrid 
\appendix  

\setcounter{equation}{0}
\setcounter{figure}{0}
\setcounter{table}{0}
\setcounter{page}{1}

\renewcommand{\theequation}{S\arabic{equation}}
\renewcommand{\thefigure}{S\arabic{figure}}
\renewcommand{\thetable}{S\arabic{table}}

\input{supplementary.tex}

\end{document}

%% file: supplementary.tex
\begin{center}
    \textbf{\large Supplemental Material: Simplicity of confinement in
      SU(3) Yang-Mills theory} \\
    \vspace{15pt}
    
    Xavier Crean,$^{1}$ Jeffrey Giansiracusa,$^{2}$ and Biagio Lucini$^{3}$ \\
    \vspace{10pt}
    
    \small
    $^{1}$\textit{Department of Mathematics, Faculty of Science and Engineering, \\ Swansea University, Fabian Way, Swansea, SA1 8EN, UK} \\
    $^{2}$\textit{Department of Mathematical Sciences, Durham University, \\ Upper Mountjoy Campus, Durham, DH1 3LE, UK} \\
    $^{3}$\textit{School of Mathematical Sciences, Queen Mary University of London, \\ Mile End Road, London, E1 4NS, UK}

    \vspace{20pt}
    
  \end{center}

\section{Topological Data Analysis for Lattice Gauge Theories}

Topology is about quantifying the aspects of shapes that are invariant under stretching and deforming.  Topological data analysis (TDA) emerged from this domain as a set of computational tools that produce numerical invariants of the shape of objects arising from data.  These numerical invariants can provide insight into the shape of the distribution underlying a set of samples.  We take a different approach here, using TDA tools to produce non-local observables for lattice gauge theories that are sensitive to large scale structures appearing in configurations.

In algebraic topology, one frequently represents geometric objects with simplicial complexes.  However, since we are working with lattice gauge theory on cubical lattices, it is more convenient to represent geometric objects as cubical complexes, which are simply objects built by gluing cubes together along their boundaries.  

One of the fundamental algebraic invariants of topology is \emph{homology}, which is about quantifying the holes and voids of a shape.  Holes come in different types. A point missing from the plane is a 1-dimensional hole in the
sense that it can be captured by a circle, which is a 1-dimensional manifold. However, a circle
cannot capture a point missing from 3-space because it could always slip off over the top or bottom;
this hole must be wrapped by a sphere, which is a 2-dimensional manifold, so this is a 2-dimensional hole.  Homology formalises this idea.  The input is a simplicial or cubical complex $X$.  The output is a vector space $H_n(X)$ for each natural number $n$.  The dimension of $H_n(X)$ represents the count of $n$-dimensional holes.
Here are some useful properties of homology.
\begin{enumerate}
\item If $X$ is a point (or contractible), then $H_0(X)$ is 1-dimensional, and $H_n(X) = {0}$ (the trivial vector space) for all $n > 0$.
\item In general, $H_0(X)$ has dimension equal to the number of connected components of X.
\item Disjoint union $X\cup Y$ corresponds to direct sum:  $H_n(X\cup Y) \cong H_n(X) \oplus H_n(Y)$.
\item $H_n(X)={0}$ for all $n$ larger than the dimension of $X$.
\item If $X$ is a $d$-dimensional closed and orientable manifold, then $H_d(X)$ is 1-dimensional.
\item If $X$ and $Y$ are homotopy equivalent, then $H_i(X) \cong H_i(Y)$ for all $i$.
\end{enumerate}

In some situations, having a vector space is more information than just having its dimension.  However, in this paper we will only make use of the dimensions.  The dimension of $H_i(X)$ is known as the $i$th Betti number of $X$, and it is denoted $b_i$.

\subsection{Cubical complexes}

An $n$-dimensional cube is simply a Cartesian product of $n$ copies of the unit interval $[0,1]$.  More systematically, given a finite set $A$, the associated cube $C(A)$ is $[0,1]^A$.  Let $\left(x_a\right)_{a\in A}$ be coordinates on the cube. A \emph{face}  of this cube is a subspace where some collection of coordinates $x_a$ are either 0 or 1. Each face is canonically linearly homeomorphic to a cube of lower dimension.  

An orientation of a cube $C(A)$ corresponds to a choice of ordering of the set $A$ up to even permutations.  Since a cube is a manifold with boundary and corners, an orientation of $C(A)$ induces an orientation of each codimension 1 face. In terms of orderings, the induced orientation of the face at $x_a = 0$ is given by representing the orientation as an ordering of $A$ with $a$ last and then restricting this to an ordering of $A\smallsetminus \{a\}$, and the induced orientation of the opposite face at $x_a = 1$ is given by the opposite of this.

A \emph{cubical complex} is a topological space $X$ together with a collection of maps $\{f_i: C_i \to X\}$ of cubes $C_i$ into $X$, satisfying the following conditions:
\begin{enumerate}
\item Each $f_i$ is a homeomorphism onto its image.
\item The union of the images is all of $X$.
\item If the images of $f_i$ and $f_j$ have non-empty intersection $K$, then the composition of $f_i$ (restricted to the preimage of $K$) followed by $f_j^{-1}$ is a linear homeomorphism of a face of $C_i$ onto a face of $C_j$.
\end{enumerate}
A \emph{subcomplex}  is a subspace of $X$ that is the union of the images of a subset of the cubes.

The example we are concerned with here is the cubical complex corresponding to a dual lattice $\Lambda^*$ for a lattice discretisation of spacetime $\Lambda$ and subcomplexes corresponding to a collection of links.

\subsection{Chain complexes and their homology}
A \emph{chain complex} is a sequence of vector spaces and linear maps 
\[
\dots \stackrel{\partial_{d+1}}{\longrightarrow} V_d \stackrel{\partial_d}{\longrightarrow} V_{d-1} \stackrel{\partial_{d-1}}{\longrightarrow} V_{d-2} \longrightarrow \dots
\]
such that the compositions $\partial_d \circ \partial_{d+1}$ are all equal to 0. 

Given a cubical complex $(X, \{f_i: C_i \to X\})$, which we abbreviate simply as $X$, one obtains a chain complex as follows.   The degree $d$ space $V_d$ is spanned by the pairs 
\[(
C_i \text{ an $d$-cube of $X$},\: o \text{ an orientation on $C_d$})
\] 
modulo the relation that $-(C, o) = (C, -o)$.  The boundary map $\partial_d$ sends $(C,o)$ to the sum of the codimension 1 faces of $C$ with their induced orientations.

Note that if all vector spaces are defined over the field $\mathbb{Z}/2$, then one doesn't need to keep track of orientations and signs, as $(C, o) = (C,-o)$.

\subsection{Homology}

Elements in the kernel of $\partial_d$ are called \emph{cycles} in degree $d$, and the subspace of cycles is written $Z_d$. Elements in the image of $\partial_{i+d}$ are called \emph{boundaries} of degree $d$, and the space of boundaries is $B_d$.  Every boundary is a cycle, so $B_d \subset Z_d$, but the converse is not necessarily true.  The homology $H_d$ is defined to be the quotient vector space $Z_d / B_d$.  It measures the extent to which there are cycles that are not boundaries.

The above algebraic definition may not appear all that intuitive, but it has the advantage that it can easily be turned into an algorithm to produce a basis for the homology in any given degree and hence determine the Betti numbers.

\subsection{Graphs}

A graph (without edges that start and end at the same vertex) is an example of a cubical complex.  In fact, these are the only kind of cubical complexes that we need to consider in this paper.

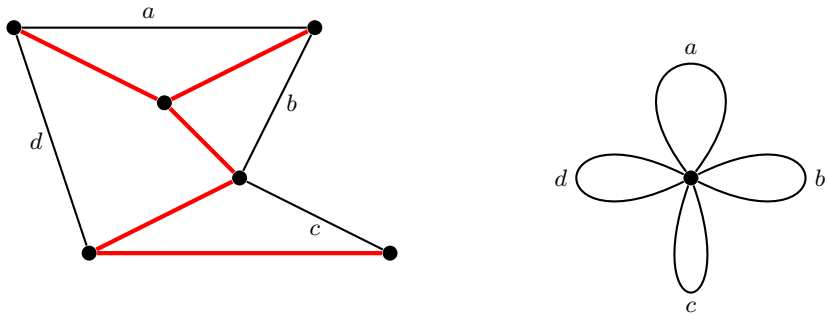
\begin{figure}
\begin{tikzpicture}[scale=1]

\tikzset{
  vertex/.style={circle, fill, inner sep=2pt},
  edge/.style={thick},
  tree/.style={ultra thick, red}
}


\node[vertex] (v1) at (0,2) {};
\node[vertex] (v2) at (2,1) {};
\node[vertex] (v3) at (4,2) {};
\node[vertex] (v4) at (3,0) {};
\node[vertex] (v5) at (1,-1) {};
\node[vertex] (v6) at (5,-1) {};

\draw[tree] (v1) -- (v2);
\draw[tree] (v2) -- (v3);
\draw[tree] (v2) -- (v4);
\draw[tree] (v4) -- (v5);
\draw[tree] (v5) -- (v6);

\draw[edge] (v1) -- node[above left] {$a$} (v3);
\draw[edge] (v4) -- node[right] {$b$} (v3);
\draw[edge] (v4) -- node[below] {$c$} (v6);
\draw[edge] (v1) -- node[left] {$d$} (v5);


\begin{scope}[xshift=9cm]

\node[vertex] (o) at (0,0) {};

\draw[edge] (o) .. controls (-1.5,2) and (1.5,2) .. node[above] {$a$} (o);
\draw[edge] (o) .. controls (2,1) and (2,-1) .. node[right] {$b$} (o);
\draw[edge] (o) .. controls (0.7,-2) and (-0.7,-2) .. node[below] {$c$} (o);
\draw[edge] (o) .. controls (-2,-1) and (-2,1) .. node[left] {$d$} (o);

\end{scope}

\end{tikzpicture}

\caption{Left: A graph $G$ and a spanning tree $T$ shown in red.  Right: the graph $G/T$ resulting from contracting $T$ to a single vertex.}
\label{fig:graph-and-spanning-tree}
\end{figure}

If $X$ is a graph, then the homology is nontrivial only in degrees 0 and 1.
The degree 0 part tells us the number of connected components, which is the
\emph{zeroth Betti number}, $b_0$.  If $G$ is a connected graph and $T \subset G$ is
a spanning tree (a subgraph that is a tree with the property that adding any
additional edge of $G$ to $T$ results in it no longer being a tree), then
contracting all the edges of $T$ results in a single vertex $v$ and a
collection of loop edges that start and end at $v$, one for each edge in the
complement of $T$.  See Fig.~\ref{fig:graph-and-spanning-tree}.  Since the
spanning tree $T$ is a tree, and hence contractible, the original graph $G$
is homotopy equivalent to the result of contracting $T$.  

One can show that the \emph{first Betti number}, $b_1$, (the dimension of $H_1(G)$) is equal to the number of edges in the complement of a spanning tree, which is the number of loops in $G/T$.

\section{\label{sec:monte-carlo-simulations}Monte Carlo simulations}
\noindent
In this study, we draw a sample of SU(3) lattice gauge field configurations using MCMC importance sampling methods, namely the heat-bath and overrelax algorithms \cite{Cabibbo:1982zn}. We define one composite update as $1$ heat-bath update followed by $4$ overrelax updates. Note that importance sampling reliably approximates the system's Boltzmann distribution once the Markov chain has been successfully equilibrated; thus, for every simulation, we discard the first $10,000$ composite updates. Further, in order to minimise the autocorrelation between successive recorded configurations, we separate each recording by $2,000$ composite updates. For each respective lattice size, we select an appropriate range of $\beta$ values to study that are strategically located to cover the critical region of the deconfinement transition; this has been previously specified in the literature, e.g., in Ref.~\cite{Lucini:2003zr}. For each $\beta$ value in our selected range, we record a sample of $N_{\text{meas}}$ configurations. Using our given computational allocation, we have been able to record $N_{\text{meas}} = 600$ configurations for lattices with $N_{t} = 4, 6$ and $N_{\text{meas}} = 400$ configurations for lattices with $N_t = 8$. Thermal expectation values of observables, as per Eq.~\eqref{eq:thermal-expectation-value}, are then estimated as the sample mean as a function of $\beta$. We estimate the standard error in the sample mean by using the bootstrap method with $N_{\text{bs}} = 2,000$. Further details on our computational pipeline can be seen in this publication's accompanying code release in Ref.~\cite{ZENODO1}.

\section{Data analysis}
\noindent
In this work, we analyse the union of the Abelian magnetic monopole currents -- treating both species on an equal footing. We have verified that considering each individual species produces compatible results to the analysis of the union of species. Further, we observe no statistically significant difference between the two species -- this is consistent with the picture seen in the literature, e.g., Refs.~\cite{Bonati:2013bga,Cardinali:2021mfh,Crean:2025gne}.

Our aim is to analyse readily computable topological invariants of the monopole current networks called Betti numbers. By fixing an arbitrary orientation (the results are independent of this choice), we map the network of monopole currents to a directed graph $G$. We then compute $b_{0} (G) \equiv \dim H_{0} (G)$ representing the components of the graph and $b_{1} (G) \equiv \dim H_{1} (G)$ representing the number of loops in the graph. 
In our computation of $b_{k}(G)$, we leverage a highly optimised implementation of a homology computing algorithm, namely Ref.~\cite{gudhi:CubicalComplex}, designed to deal with a cubical complex. This is possible since a directed graph is a $1$-dimensional cubical complex; therefore, a cubical complex is a convenient data structure for representing graphs. See Ref.~\cite{Crean:2024nro}, for further details. 

\begin{figure}
    \centering
    \includegraphics[   width=0.45\columnwidth]{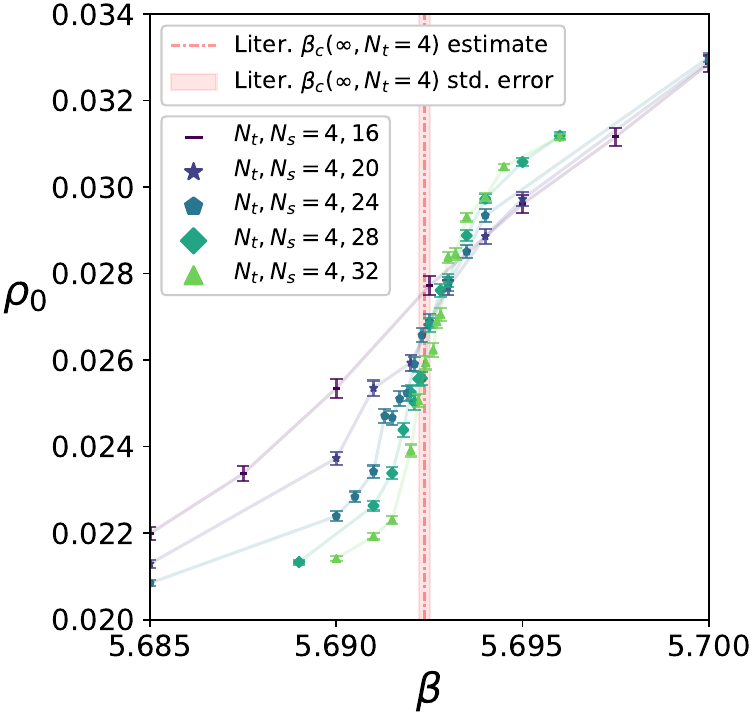}
    \includegraphics[   width=0.45\columnwidth]{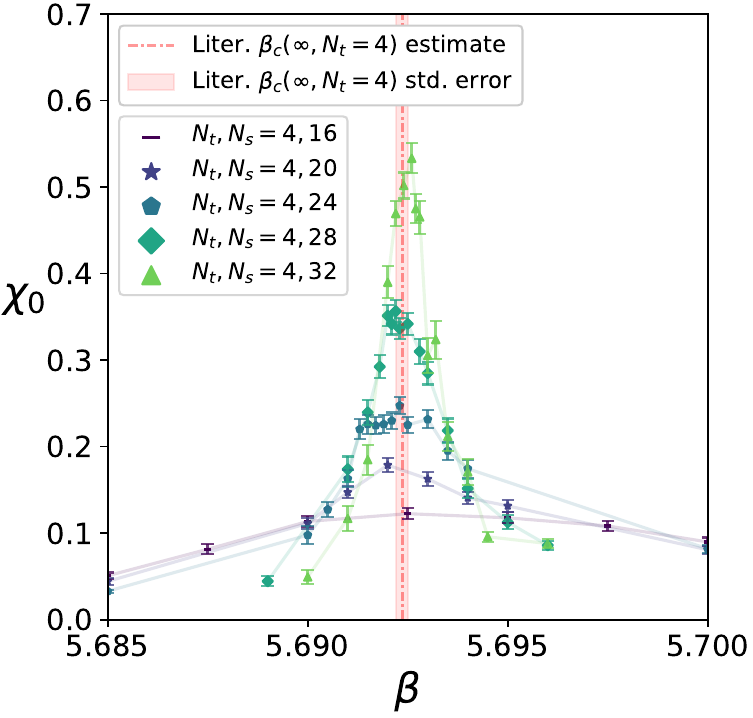}
    \caption{For $N_t = 4$, a scatter plot of $\rho_0$ and $\chi_0$ as functions of $\beta$ with translucent lines to guide the eye. The vertical line and band show respectively the central value and the statistical error for the extrapolated $\beta_{c}$ determined in Ref.~\cite{Lucini:2003zr}.}
    \label{fig:Nt=4_p0_scatter}
\end{figure}

\begin{figure}
    \centering
    \includegraphics[   width=0.45\columnwidth]{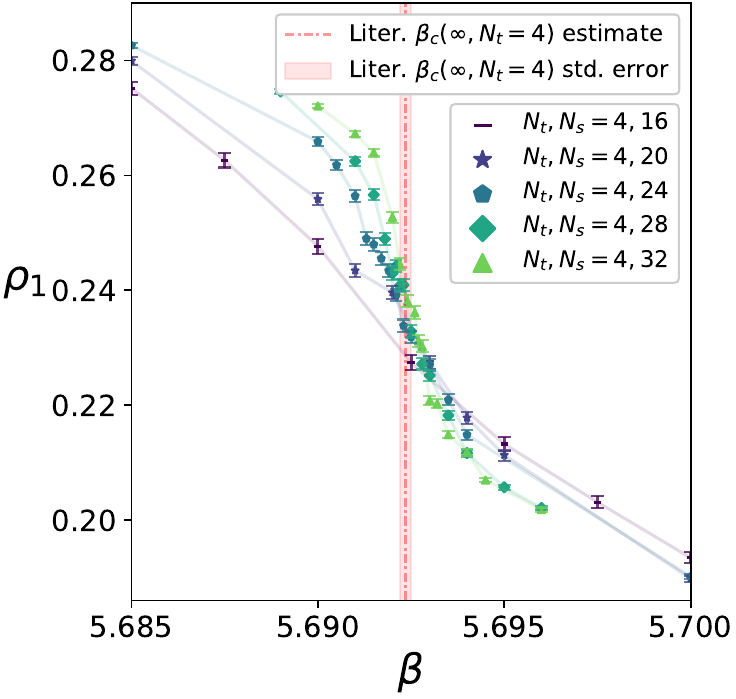}
    \includegraphics[   width=0.45\columnwidth]{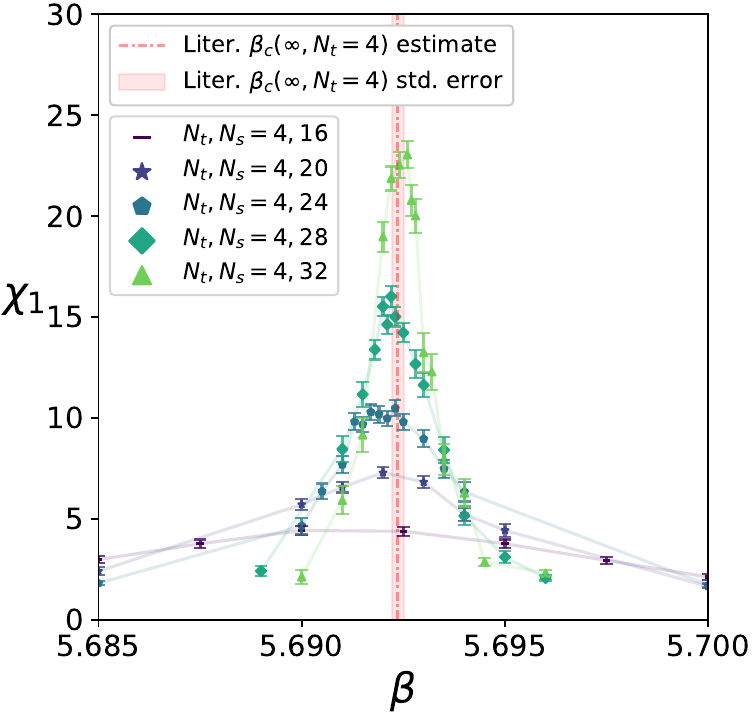}
    \caption{For $N_t = 4$, a scatter plot of $\rho_1$ and $\chi_1$ as functions of $\beta$ with translucent lines to guide the eye. The vertical line and band show respectively the central value and the statistical error for the extrapolated $\beta_{c}$ determined in Ref.~\cite{Lucini:2003zr}.}
    \label{fig:Nt=4_p1_scatter}
\end{figure}

\begin{figure}
    \centering
    \includegraphics[   width=0.45\columnwidth]{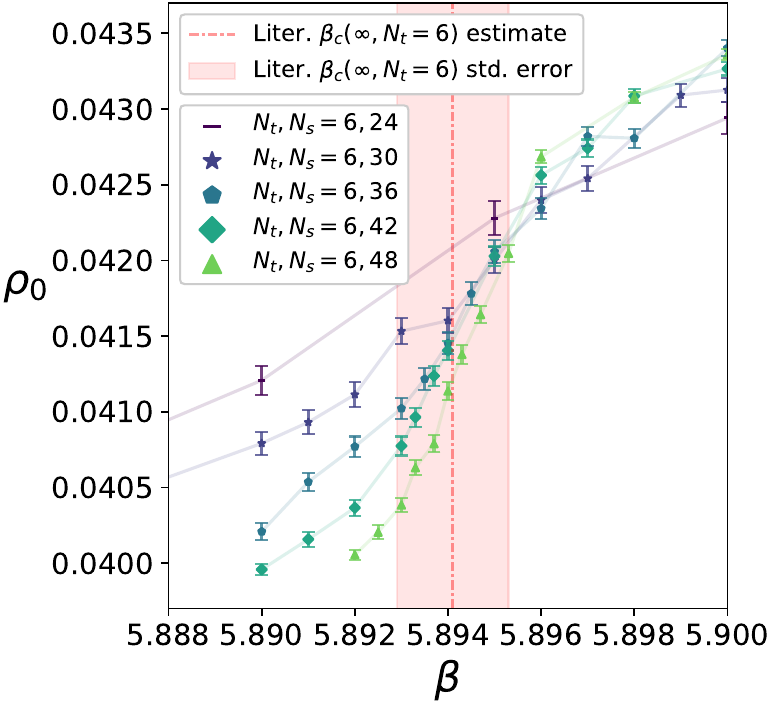}
    \includegraphics[   width=0.45\columnwidth]{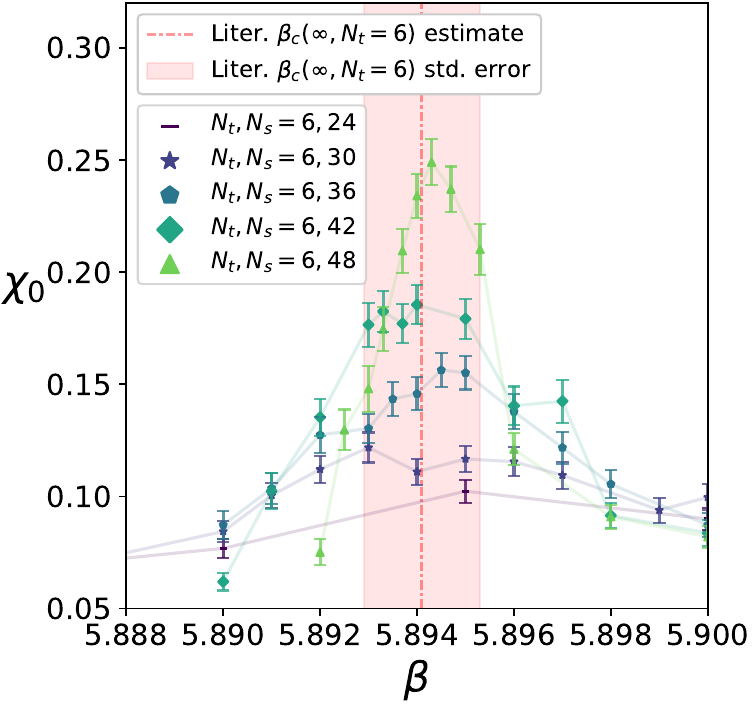}
    \caption{For $N_t = 6$, a scatter plot of $\rho_0$ and $\chi_0$ as functions of $\beta$ zoomed into the critical region with translucent lines to guide the eye. The vertical line and band show respectively the central value and the statistical error for the extrapolated $\beta_{c}$ determined in Ref.~\cite{Lucini:2003zr}.}
    \label{fig:Nt=6_p0_scatter}
\end{figure}

\begin{figure}
    \centering
    \includegraphics[   width=0.45\columnwidth]{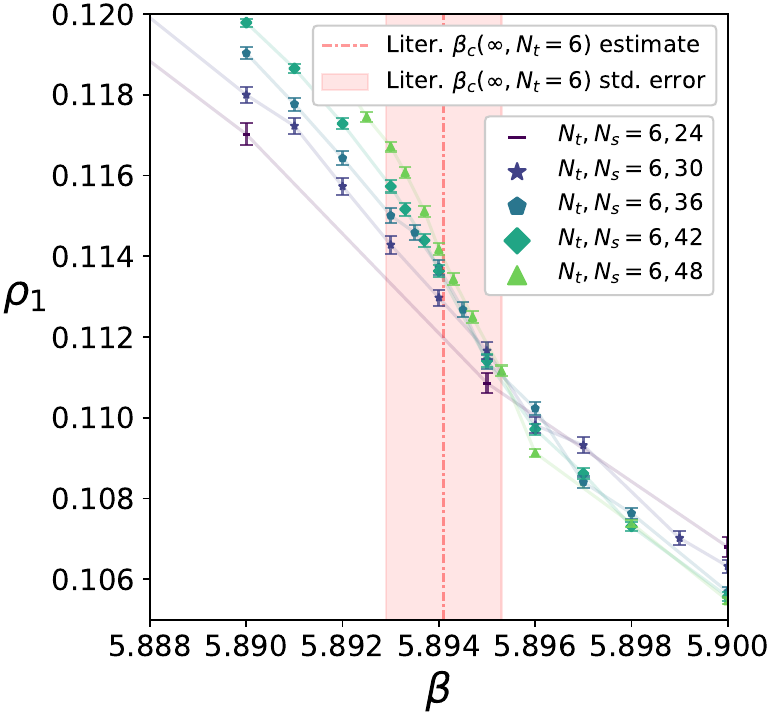}
    \includegraphics[   width=0.45\columnwidth]{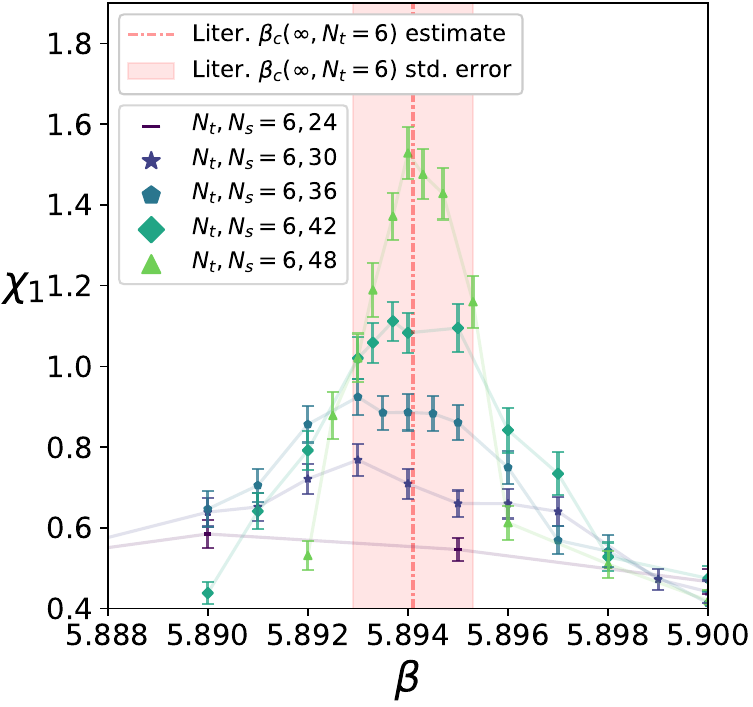}
    \caption{For $N_t = 6$, a scatter plot of $\rho_1$ and $\chi_1$ as functions of $\beta$ zoomed into the critical region with translucent lines to guide the eye. The vertical line and band show respectively the central value and the statistical error for the extrapolated $\beta_{c}$ determined in Ref.~\cite{Lucini:2003zr}.}
    \label{fig:Nt=6_p1_scatter}
\end{figure}

\begin{figure}
    \centering
    \includegraphics[   width=0.45\columnwidth]{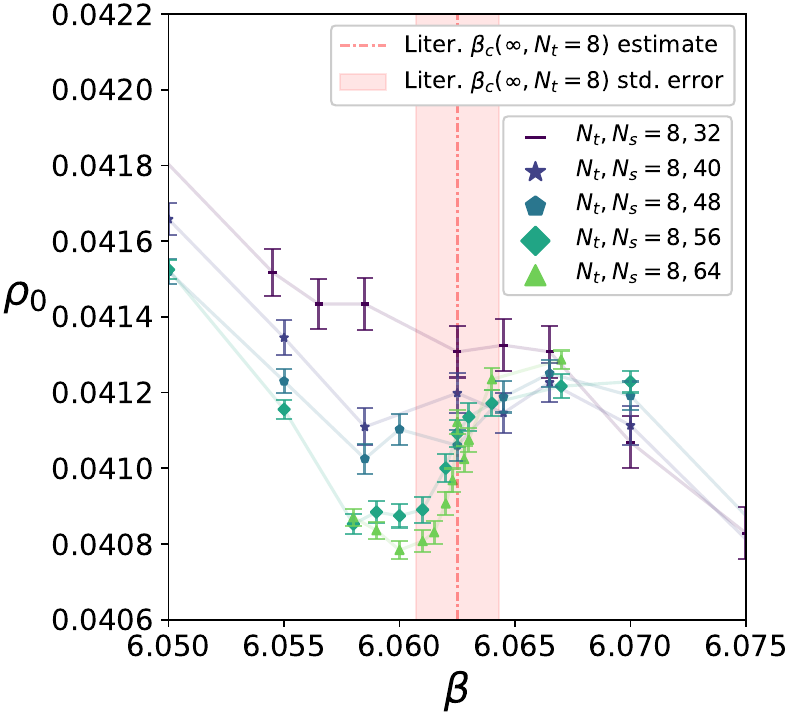}
    \includegraphics[   width=0.45\columnwidth]{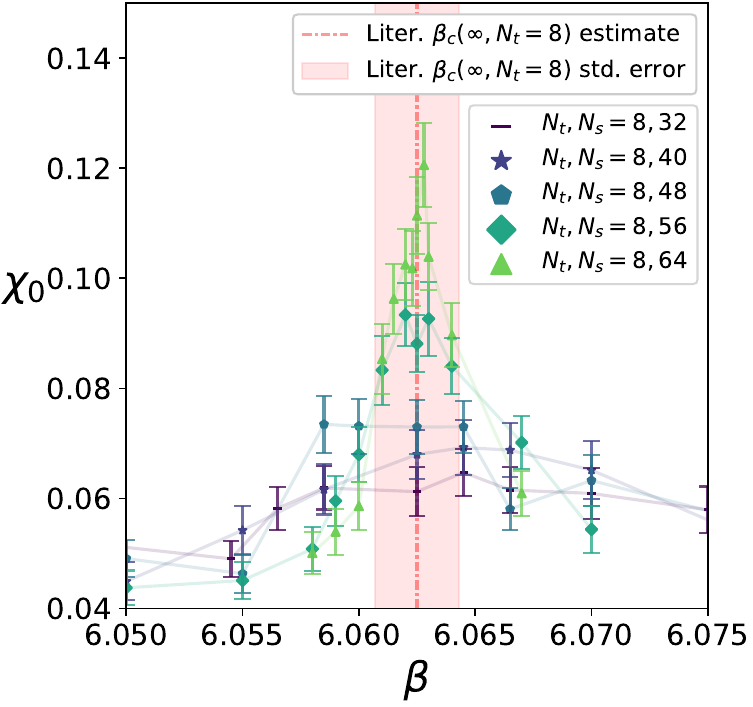}
    \caption{For $N_t = 8$, a scatter plot of $\rho_0$ and $\chi_0$ as functions of $\beta$ zoomed into the critical region with translucent lines to guide the eye. The vertical line and band show respectively the central value and the statistical error for the extrapolated $\beta_{c}$ determined in Ref.~\cite{Lucini:2003zr}.}
    \label{fig:Nt=8_p0_scatter}
\end{figure}

\begin{figure}
    \centering
    \includegraphics[   width=0.45\columnwidth]{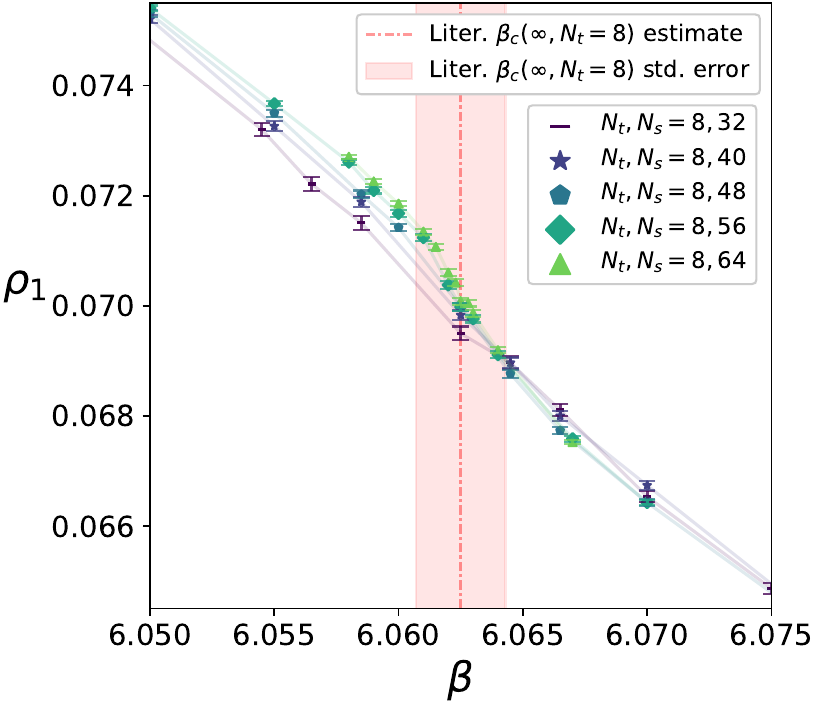}
    \includegraphics[   width=0.45\columnwidth]{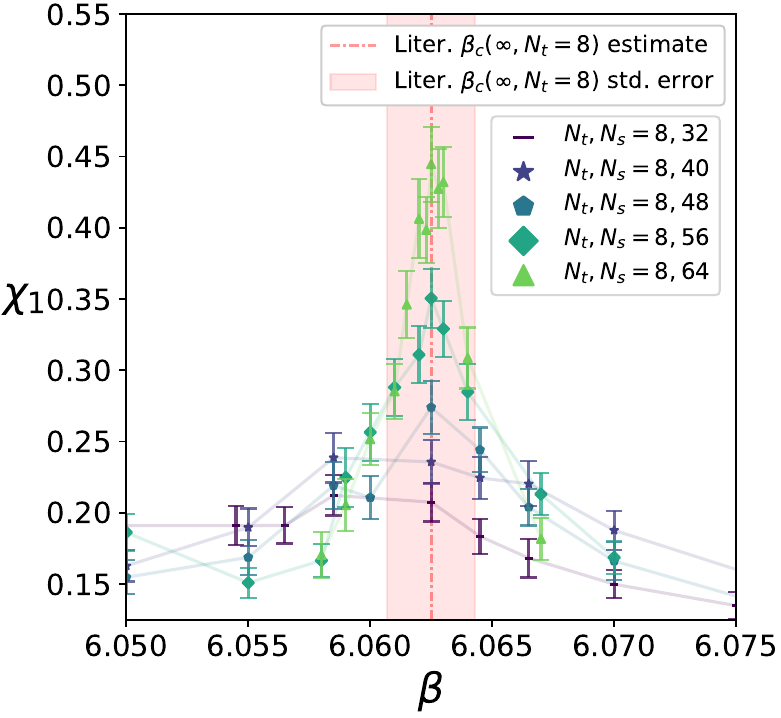}
    \caption{For $N_t = 8$, a scatter plot of $\rho_1$ and $\chi_1$ as functions of $\beta$ zoomed into the critical region with translucent lines to guide the eye. The vertical line and band show respectively the central value and the statistical error for the extrapolated $\beta_{c}$ determined in Ref.~\cite{Lucini:2003zr}.}
    \label{fig:Nt=8_p1_scatter}
\end{figure}

Our results for $\rho_i (\beta)$ and $\chi_i (\beta)$ (as defined in the main body) at $N_t = 4, 6, 8$ are plotted in Figs.~\ref{fig:Nt=4_p0_scatter}, \ref{fig:Nt=4_p1_scatter}, \ref{fig:Nt=6_p0_scatter}, \ref{fig:Nt=6_p1_scatter}, \ref{fig:Nt=8_p0_scatter} and \ref{fig:Nt=8_p1_scatter} respectively. A suitable range of $\beta$-values covering the critical region have been chosen with error bars computed using the bootstrap method with $N_{\text{bs}} = 2,000$. One can see that the susceptibilities $\chi_{i}$ peak at the critical value with peaks becoming larger and more concentrated as the spatial volume increases as expected for an extensive observable at the critical value of a first order phase transition.



\begin{figure}
    \centering
    \includegraphics[   width=0.45\columnwidth]{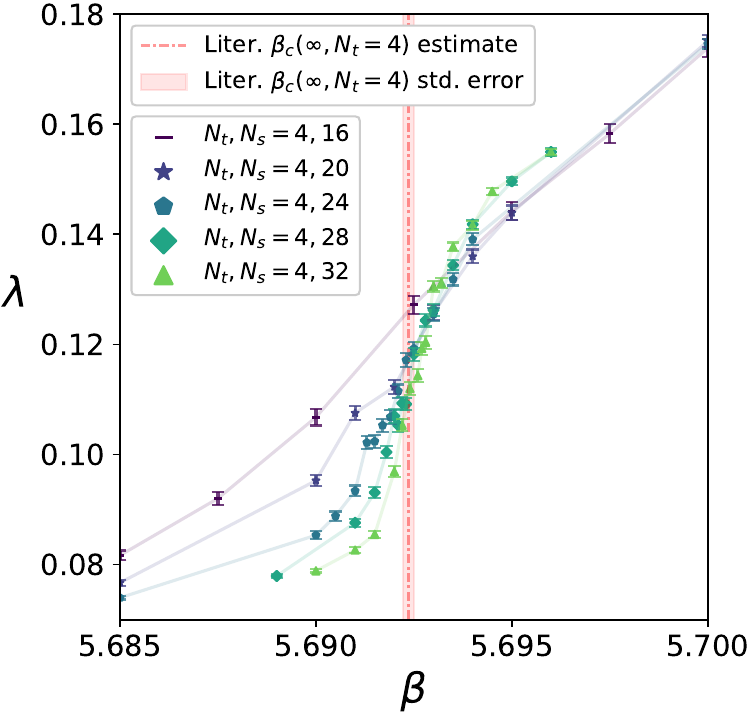}
    \includegraphics[   width=0.45\columnwidth]{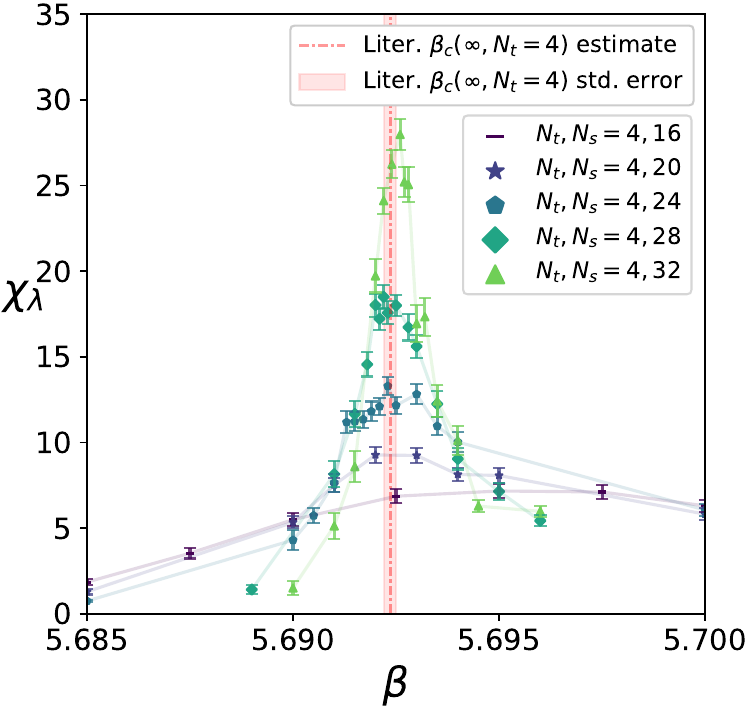}
    \caption{For $N_t = 4$, a scatter plot of $\lambda$ and $\chi_\lambda$ as functions of $\beta$ with translucent lines to guide the eye. The vertical line and band show respectively the central value and the statistical error for the extrapolated $\beta_{c}$ determined in Ref.~\cite{Lucini:2003zr}.}
    \label{fig:Nt=4_scatter}
\end{figure}

\begin{figure}
    \centering
    \includegraphics[   width=0.45\columnwidth]{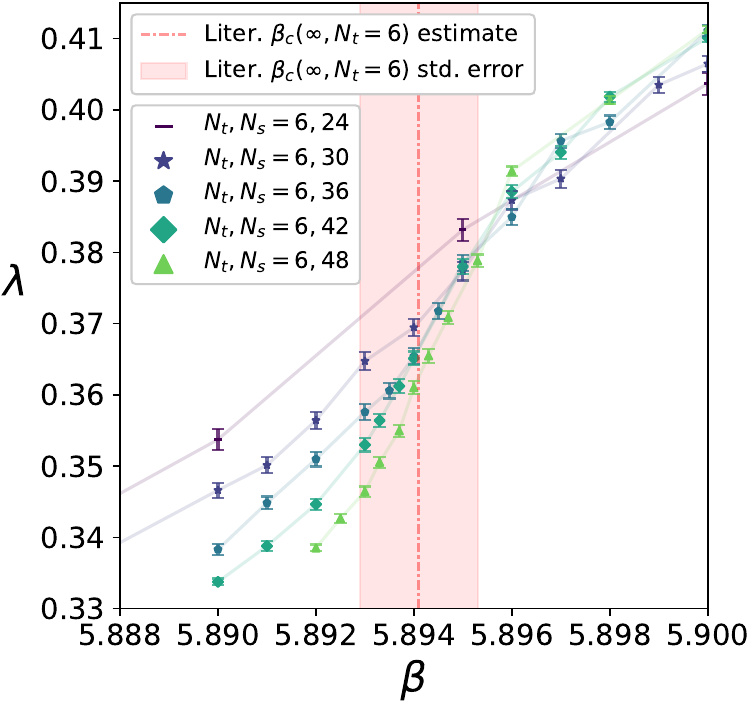}
    \includegraphics[   width=0.45\columnwidth]{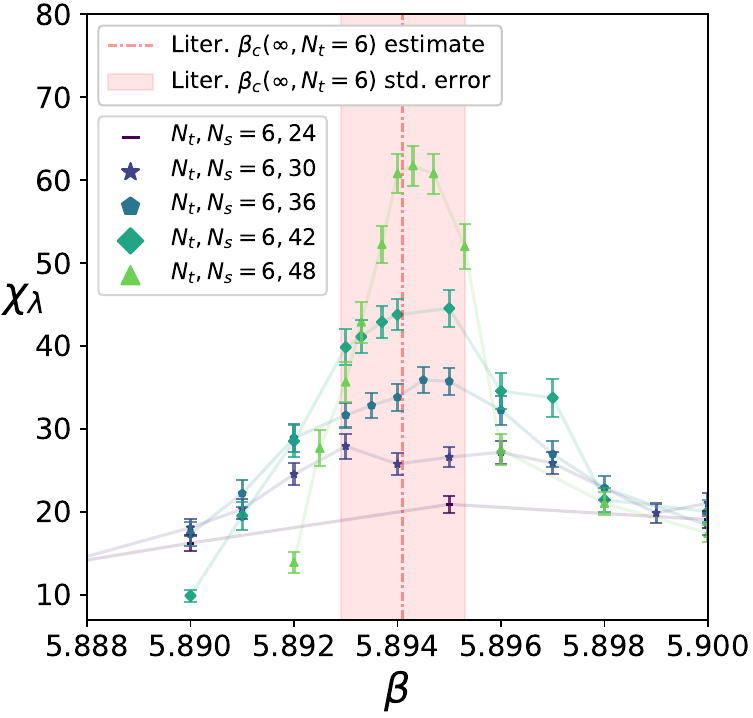}
    \caption{For $N_t = 6$, a scatter plot of $\lambda$ and $\chi_\lambda$ as functions of $\beta$ with translucent lines to guide the eye. The vertical line and band show respectively the central value and the statistical error for the extrapolated $\beta_{c}$ determined in Ref.~\cite{Lucini:2003zr}.}
    \label{fig:Nt=6_scatter}
\end{figure}

\begin{figure}
    \centering
    \includegraphics[   width=0.45\columnwidth]{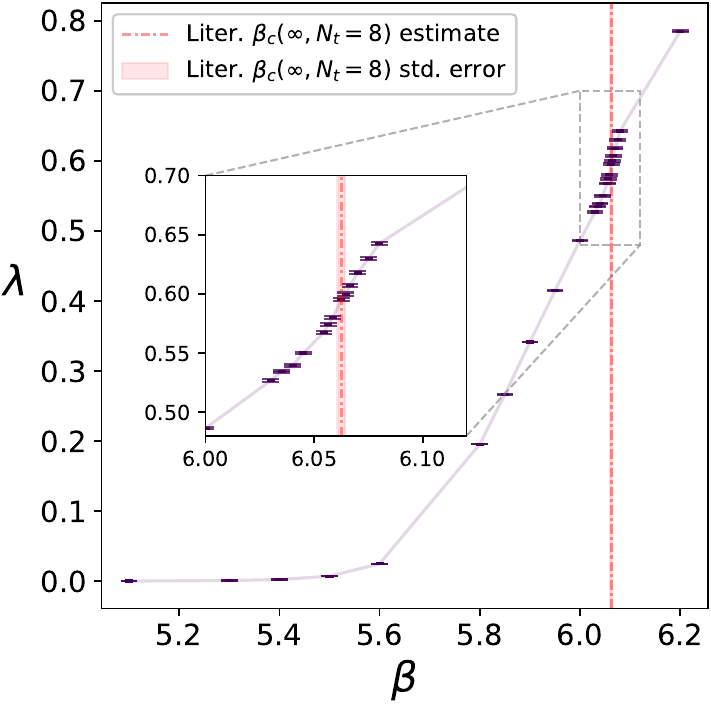}
    \includegraphics[   width=0.45\columnwidth]{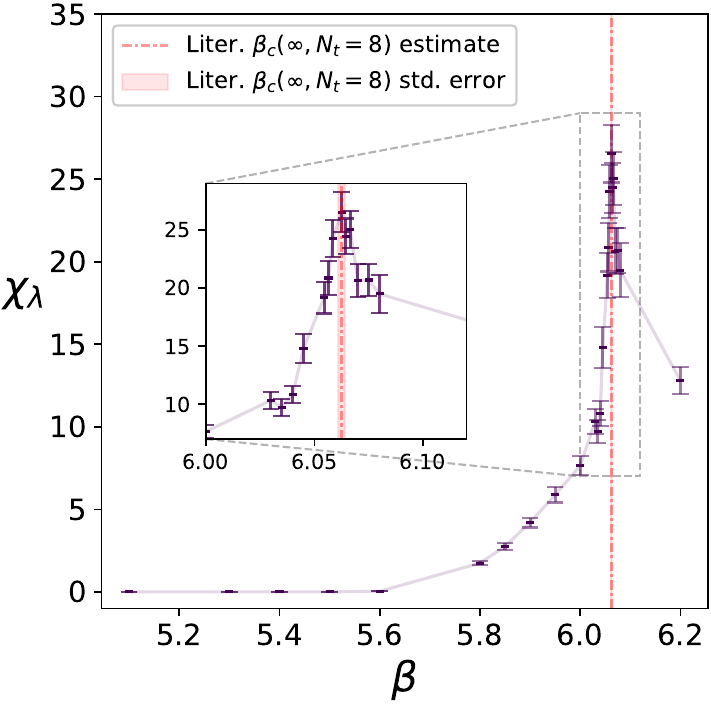}
    \caption{For $N_t , N_s = 8, 32$, a scatter plot of $\lambda$ and $\chi_\lambda$ as functions of $\beta$ over a wider range of $\beta$ -- with inset plot zoomed into the critical region. Translucent lines are included to guide the eye. One can see that $\lambda \in [0,1]$ as expected. The red vertical line and band show respectively the central value and the statistical error for the extrapolated $\beta_{c}$ determined in Ref.~\cite{Lucini:2003zr}.}
    \label{fig:Nt=8_scatter_wide}
\end{figure}

Similarly, our results for the simplicity $\lambda  (\beta)$ and $\chi_{\lambda} (\beta)$ for $N_t = 4, 6, 8$ are plotted in Figs.~\ref{fig:Nt=4_scatter}, \ref{fig:Nt=6_scatter} and \ref{fig:Nt=8_scatter} respectively. One can see that the susceptibility $\chi_{\lambda}$ peaks at the critical value with peaks becoming larger and more concentrated as the spatial volume increases. In Fig.~\ref{fig:Nt=8_scatter_wide}, we plot a wider range of $\beta$-values for $N_t , N_s = 8 , 32$ so that the global trend is made clear for finite $N_s$ size.


\subsection{Reweighting via density of states estimation}
\noindent
Our aim is to estimate the location of the peaks of $\chi_{i}$ ($i = 0,1,\lambda$) so that we may extrapolate to the infinite volume limit $N_{s} \to \infty$. We use multiple histogram reweighting to extract a more precise location of the peak by estimating $\chi_{i}(\beta)$ at a high resolution of $\beta$-values covering the peak of our simulated values. This procedure involves learning the density of states of the system $\rho(E)$ where the partition function may be expressed
\begin{equation}
    Z_{\beta} = \sum_{E} \rho(E) e^{-\beta E}.
\end{equation} 
In practice, following Ref.~\cite{Ferrenberg:1989ui}, this is achieved implicitly by solving via iteration the equation
\begin{equation}\label{eq:self-consistency}
    Z_{\beta} = \sum_{i=1}^{R} \sum_{a=1}^{N_{i}} \frac{g_{i}^{-1} e^{-\beta E_{i}^{a}}}{\sum_{j=1}^{R} N_{j} g_{j}^{-1} e^{-\beta_{j} E_{i}^{a} - \log Z_{\beta_{j}}} }
\end{equation}
where 
\begin{itemize}
    \item $R$ is the number of simulations each respectively conducted at $\beta_{i}$ with sample size $N_{i}$,
    \item $E_{i}^{a}$ is the energy measurement (in our case action) of the $a$-th recorded configuration in a simulation
    \item and $g_{i} = 1 + 2 \tau_{i}$ is a coefficient measuring the autocorrelation time between recorded configurations.
\end{itemize}
We may then estimate the expectation value of an observable $O$ using 
\begin{equation}
    \langle O \rangle_{\beta} \approx \sum_{i=1}^{R} \sum_{a=1}^{N_{i}} \frac{O_{i}^{a} g_{i}^{-1} e^{-\beta E_{i}^{a} + \log Z_{\beta}}}{\sum_{j=1}^{R} N_{j} g_{j}^{-1} e^{-\beta_{j} E_{i}^{a} - \log Z_{\beta_{j}}} }.
\end{equation}
Note that it can be a challenge to keep exponentials of the form $\exp\{ - (\Delta \beta) S\}$ in a stable numerical range (avoiding underflow to 0) especially as the action $S$ scales with lattice spacetime volume $N_{t}N_{s}^{3}$. Our algorithmic implementation in Ref.~\cite{ZENODO1} uses long double precision and recentring to avoid numerical underflow.

In order to estimate the systematic error in the reweighting procedure, we use the bootstrap method with $N_{\text{bs}} = 2,000$. More specifically, following Ref.~\cite{Ferrenberg:1989ui}, given a simulation at $\beta_{i}$ of sample size $N_{i}$, we estimate the autocorrelation $\tau_{i}$ between recorded configurations in the sample. Approximating $g_{i} \approx \tau_{i}$, we then randomly subsample to produce an effective sample of decorrelated configurations of size 
\begin{equation}\label{eq:effective-decorrelated-configurations}
    N_{i}^{\text{eff}} = \lfloor N_{i} / \tau_{i} \rfloor \ ,
\end{equation}
where $\lfloor \cdot \rfloor$ is the floor function. To minimise autocorrelation, we use the RANLUX random number generator defined in Ref.~\cite{Luscher:1993dy}. Since configurations in this bootstrap sample are effectively decorrelated, this allows us to iteratively solve Eq.~\eqref{eq:self-consistency} with $g_{i} = 1$. We take a convergence tolerance of $10^{-14}$ (smaller tolerances show compatible convergence). We thus have a resultant bootstrap distribution consisting of $2,000$ reweighting curves.

We extract $\beta_{c} = \text{argmax}\{ \chi_{i}(\beta) \}$, the location of the peak of each curve, for every sample in the bootstrap distribution. We then take the mean as the estimate and standard deviation as the standard error.

\section{\label{sec:fitting-of-numerical-results}Fitting of numerical results}
\noindent
In order to extrapolate the peaks of our observables $\beta_{c} (N_{s}, N_{t})$, referred to as pseudo-critical values, to the thermodynamic limit $N_{s} \to \infty$, we fit the data using a polynomial ansatz in inverse powers of the spatial volume
\begin{equation}\label{eq:polynomial-ansatz}
    \beta_{c}(N_{s}, N_{t}) = \beta_{c}(N_{t}) + \sum_{k = 1}^{k_{\text{max}}} \alpha_{k} (N_{s}^{3})^{-k}
\end{equation}
where $k_{\text{max}} < \infty$ defines the degree of the polynomial. We perform a variety of fits varying the range of the data and polynomial degree. Note that we discard models with an extremal value within the range of datapoints since only subleading non-linear corrections are valid in the thermodynamic limit. The estimate for $\beta_{c}(N_{t})$ is the intercept of each fit and we estimate the standard error again using the bootstrap method using $N_{\text{bs}} = 2,000$.

\begin{figure}
    \centering
    \includegraphics[width=0.45\columnwidth]{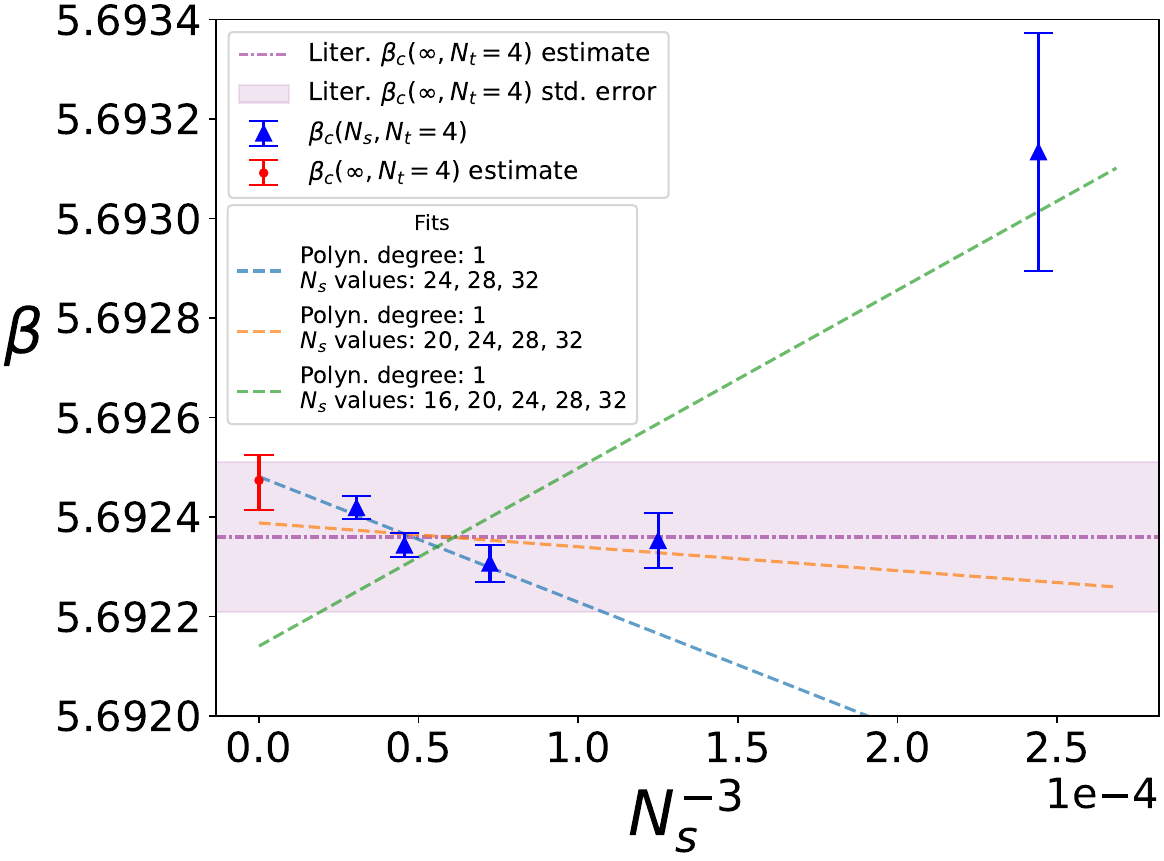} 
    \includegraphics[width=0.45\columnwidth]{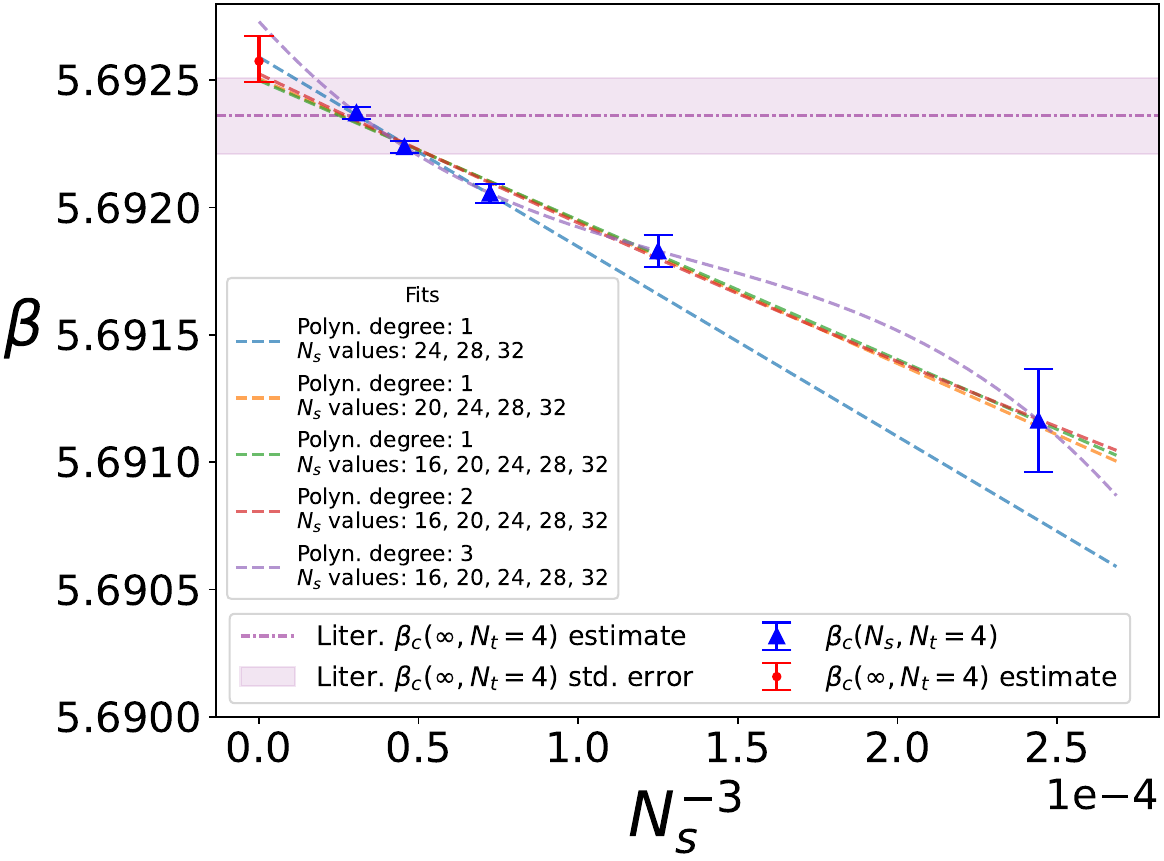}
    \caption{Finite-size scaling analysis for the observables $\rho_{0}$ (left) and $\rho_{1}$ (right) both for lattices with $N_t = 4$ and $N_s = 16,20,24,28,32$. Blue triangles represent $\beta_{c}(N_{s}, N_{t}=4)$, i.e., location of the respective peaks of the reweighted susceptibility curves, with error bars computed via bootstrapping with $N_{bs} = 2,000$. Dashed lines represent the polynomial regression fits used for the infinite volume extrapolation of these peak values. The red point is our calculated estimate of the $\beta_{c}$ in the thermodynamic limit $N_{s} \to \infty$ using the weighting procedure outlined in the body. The horizontal line and band show respectively the central value and the statistical error for the extrapolated $\beta_{c}$ determined in Ref.~\cite{Lucini:2003zr}.}
    \label{fig:fits_Nt4}
\end{figure}

\begin{figure}
    \centering
    \includegraphics[width=0.45\columnwidth]{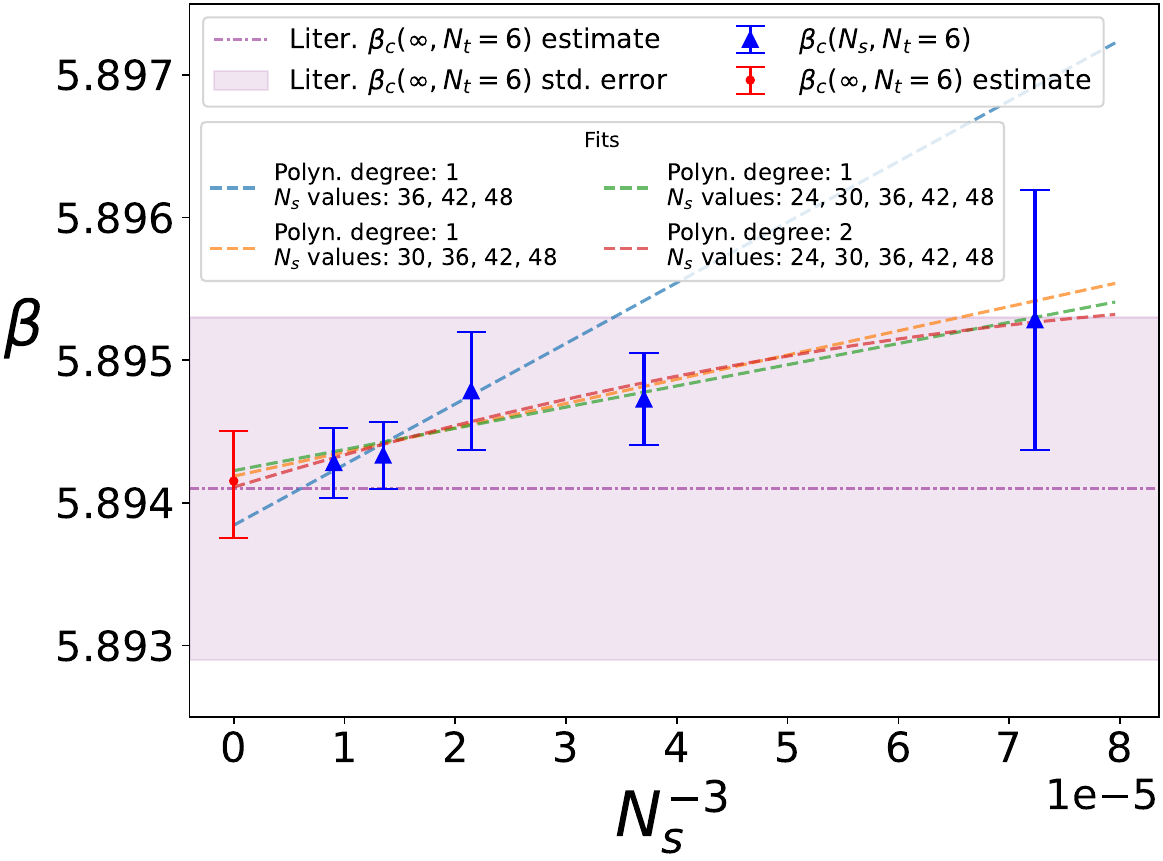} 
    \includegraphics[width=0.45\columnwidth]{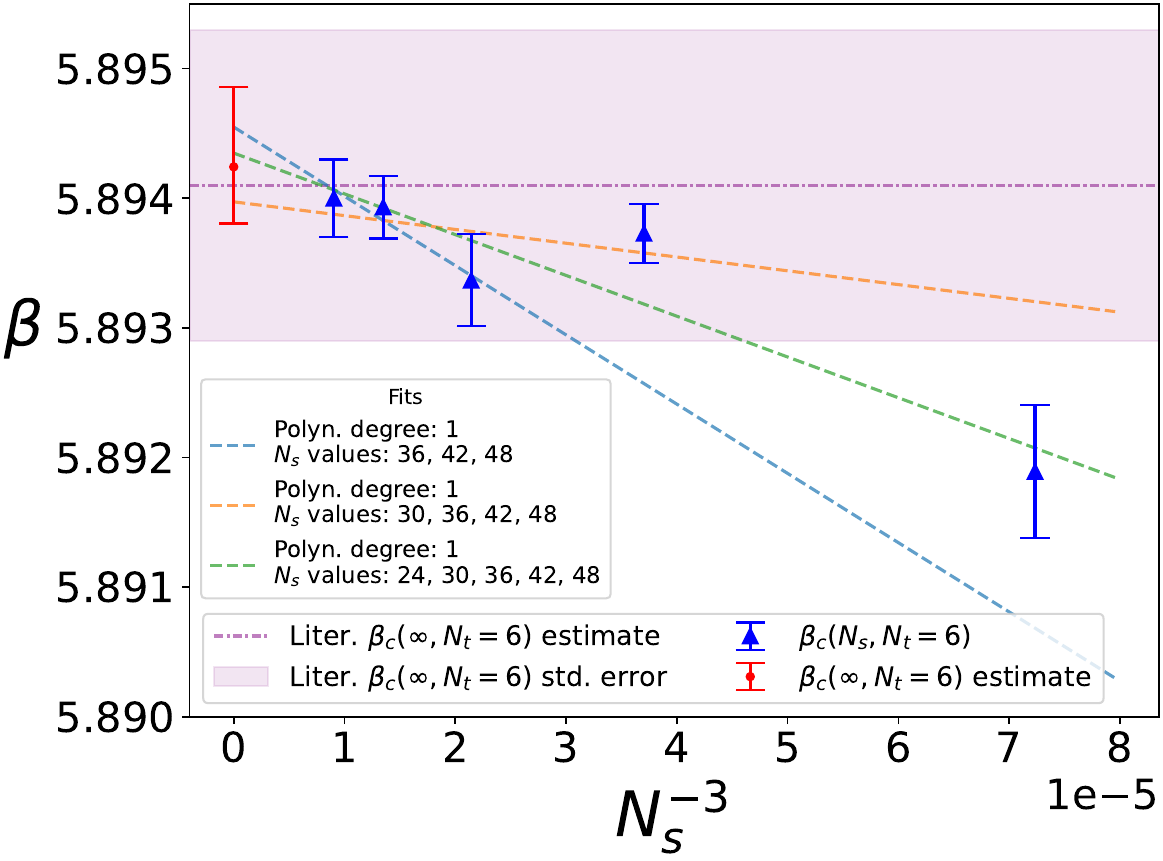}
    \caption{Finite-size scaling analysis for the observables $\rho_{0}$ (left) and $\rho_{1}$ (right) both for lattices with $N_t = 6$ and $N_s = 24,30,36,42,48$. Blue triangles represent $\beta_{c}(N_{s}, N_{t}=6)$, i.e., location of the respective peaks of the reweighted susceptibility curves, with error bars computed via bootstrapping with $N_{bs} = 2,000$. Dashed lines represent the polynomial regression fits used for the infinite volume extrapolation of these peak values. The red point is our calculated estimate of the $\beta_{c}$ in the thermodynamic limit $N_{s} \to \infty$ using the weighting procedure outlined in the body. The horizontal line and band show respectively the central value and the statistical error for the extrapolated $\beta_{c}$ determined in Ref.~\cite{Lucini:2003zr}.}
    \label{fig:fits_Nt6}
\end{figure}

\begin{figure}
    \centering
    \includegraphics[
    width=0.45\columnwidth]{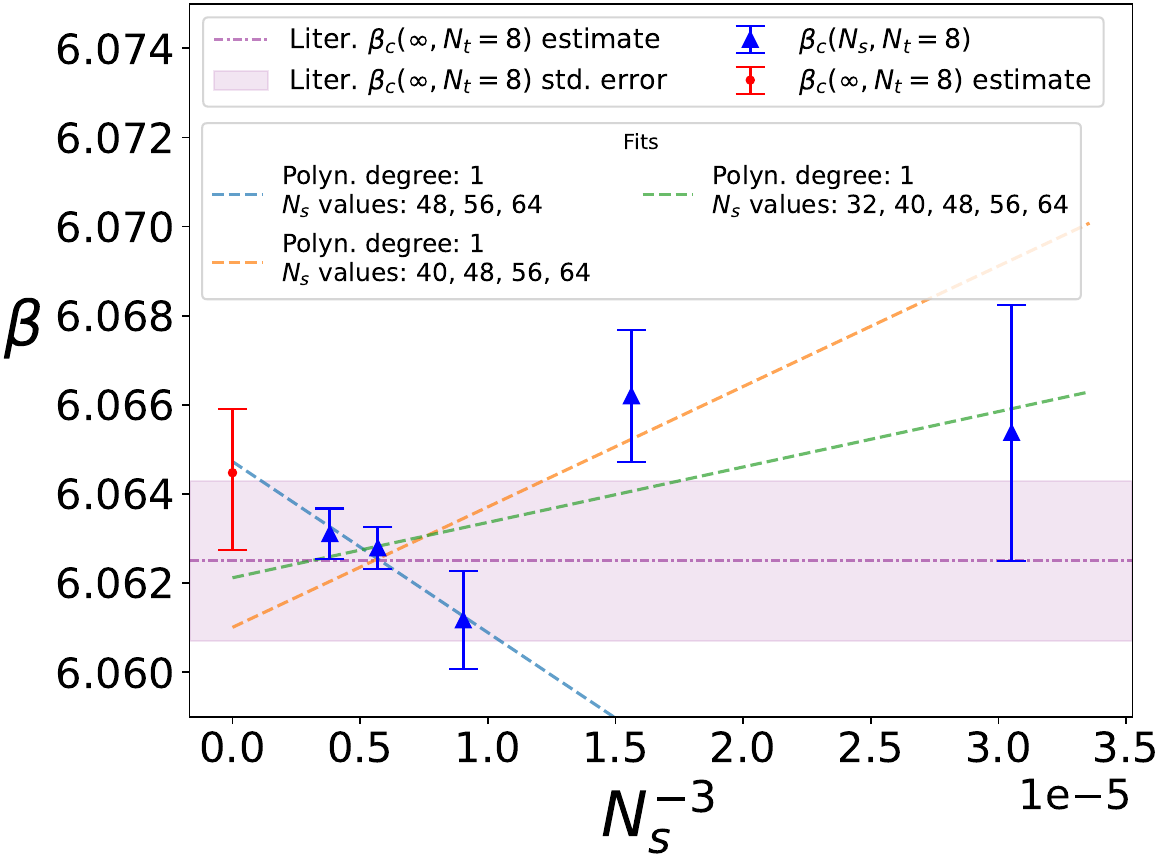} 
    \includegraphics[   width=0.45\columnwidth]{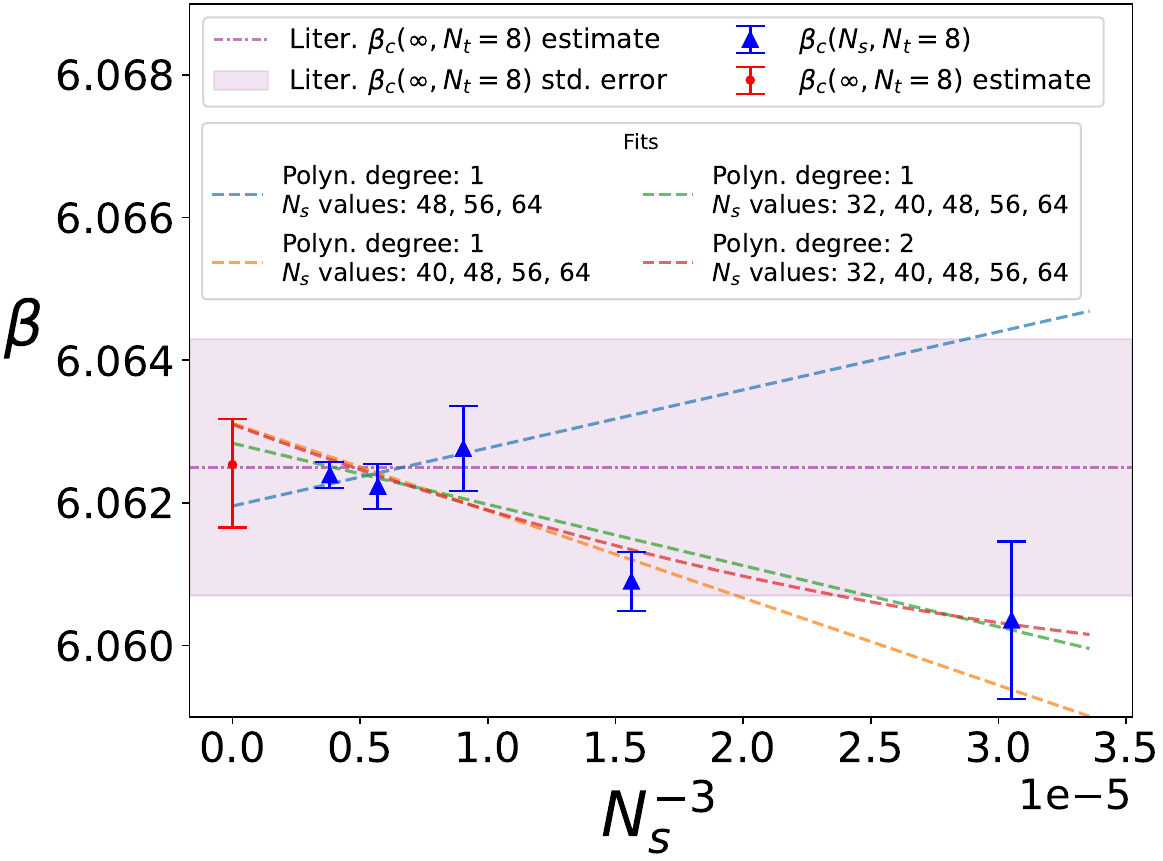}
    \caption{Finite-size scaling analysis for the observables $\rho_{0}$ (left) and $\rho_{1}$ (right) both for lattices with $N_t = 8$ and $N_s = 32,40,48,56,64$. Blue triangles represent $\beta_{c}(N_{s}, N_{t}=8)$, i.e., location of the respective peaks of the reweighted susceptibility curves, with error bars computed via bootstrapping with $N_{bs} = 2,000$. Dashed lines represent the polynomial regression fits used for the infinite volume extrapolation of these peak values. The red point is our calculated estimate of the $\beta_{c}$ in the thermodynamic limit $N_{s} \to \infty$ using the weighting procedure outlined in the body. The horizontal line and band show respectively the central value and the statistical error for the extrapolated $\beta_{c}$ determined in Ref.~\cite{Lucini:2003zr}.}
    \label{fig:fits_Nt8}
\end{figure}

Rather than select an individual fit, we produce an estimate based on an ensemble of fits. Following Ref.~\cite{Jay:2020jkz}, we use a statistic constructed from the Akaike Information Criterion (AIC) that considers goodness of fit, model complexity and degrees of freedom in the model. Roughly, the idea is to give strong weighting to models that have a high goodness of fit score, low model complexity and use all available data points. The normalised weights, interpreted as a probability, are calculated using the expression
\begin{equation}\label{eq:model-fit-weight}
    w_{i} = \frac{1}{\mathcal{N}} \exp \{ -\frac{1}{2}(\chi^{2} + 2 n_{\text{par}} - n_{\text{data}} )\}
\end{equation}
where $\chi^{2}$ is the standard chi-squared statistic, $n_{\text{par}}$ is the number of parameters in the model, $n_{\text{data}}$ is the number of data points used in the fit and $\mathcal{N}$ is the normalisation with respect to all fits. We then model the distribution by a weighted sum of Gaussian distributions with PDF
\begin{equation}\label{eq:weighted-sum-of-gaussians}
    p (x) = \sum_{i} w_i N(x; \mu_{i}, \sigma_{i})
\end{equation}
with mean $\mu_{i}$ and standard deviation $\sigma_{i}$ taken to be the statistical error in the estimate of the intercept (as estimated above via the bootstrap method). Taking the CDF of this distribution, we are able to solve numerically for the $16\%$, $50\%$, $84\%$ centiles as our lower bound, estimate and upper bound respectively. This final estimate represents weighted contributions from all fits and so we claim this is more robust than quoting estimates based on any single individual fit. Our estimates of $\beta_c$ are presented in Tab.~\ref{tab:model-fits-details}. Plots of scaling analysis for $\rho_0$ and $\rho_1$ are presented in Figs.~\ref{fig:fits_Nt4}, \ref{fig:fits_Nt6} and \ref{fig:fits_Nt8}. Plots of scaling analysis for $\lambda$ are presented in Figs.~\ref{fig:fits_ratio_Nt4}, \ref{fig:fits_ratio_Nt6} and \ref{fig:fits_ratio_Nt8}. 
\begin{table*}[]
    \centering
    \renewcommand{\arraystretch}{3.0}
    \setlength{\tabcolsep}{8pt}
    \begin{tabular}{|c|c|c|c|}
        \hline
        Observable & $N_t = 4$ & $N_t = 6$ & $N_t = 8$ \\
        \hline
        \hline
        $\rho_0$ & $5.69247^{+0.00005}_{-0.00006}$
 & $5.8942^{+0.0004}_{-0.0004}$
 & $6.0645^{+0.0014}_{-0.0017}$ \\
        \hline
        $\rho_1$ & $5.69257^{+0.00010}_{-0.00008}$ & $5.8942^{+0.0006}_{-0.0004}$ & $6.0625^{+0.0006}_{-0.0009}$ \\
        \hline
        $\lambda$ & $5.69240^{+0.00005}_{-0.00005}$ & $5.8940^{+0.0004}_{-0.0004}$ & $6.0625^{+0.0008}_{-0.0010}$ \\
        \hline
        \hline
         Literature value & $5.69236^{+0.00015}_{-0.00015}$ & $5.8941^{+0.0012}_{-0.0012}$ & $6.0625^{+0.0018}_{-0.0018}$ \\
        \hline
    \end{tabular}
    \caption{Performing a finite-size scaling analysis using the Betti number observables, simplicity and their respective susceptibilities, this table specifies our estimates for $\beta_{c}$ in the thermodynamic limit $N_{s} \to \infty$ for lattices with temporal size $N_t = 4,6,8$. Error bars are computed using the $68 \%$ confidence interval of the weighted sum of Gaussians encoding the various fits. As a comparison with the literature, the final row are results from Ref.~\cite{Lucini:2003zr}. This study used comparable MC methods to generate its configurations, which allows us to make a faithful comparison between the sensitivity (to the deconfinement phase transition) of our Abelian monopole observables with standard observables. In that study, the expectation value of the modulus of the average Polyakov loop $\langle |\bar{P}| \rangle$ was used such that the peaks of the respective susceptibilities $\chi_{P}(\beta)$ were used to give an estimate for $\beta_{c}$ as $N_{s} \to \infty$. One can see that our results are compatible with those given in the literature.}
    \label{tab:model-fits-details}
\end{table*}


%% file: refs_new.bib
@article{Aarts:2023vsf,
    author = "Aarts, Gert and others",
    title = "{Phase Transitions in Particle Physics}: {Results and Perspectives from Lattice Quantum Chromo-Dynamics}",
    eprint = "2301.04382",
    archivePrefix = "arXiv",
    primaryClass = "hep-lat",
    doi = "10.1016/j.ppnp.2023.104070",
    journal = "Prog. Part. Nucl. Phys.",
    volume = "133",
    pages = "104070",
    year = "2023"
}

@article{Glozman:2016swy,
    author = "Glozman, L. Ya.",
    title = "{$SU(2N_F)$ symmetry of QCD at high temperature and its implications}",
    eprint = "1610.00275",
    archivePrefix = "arXiv",
    primaryClass = "hep-lat",
    doi = "10.5506/APhysPolBSupp.10.583",
    journal = "Acta Phys. Polon. Supp.",
    volume = "10",
    pages = "583",
    year = "2017"
}

@article{Alexandru:2019gdm,
    author = "Alexandru, Andrei and Horv{\'a}th, Ivan",
    title = "{Possible New Phase of Thermal QCD}",
    eprint = "1906.08047",
    archivePrefix = "arXiv",
    primaryClass = "hep-lat",
    doi = "10.1103/PhysRevD.100.094507",
    journal = "Phys. Rev. D",
    volume = "100",
    number = "9",
    pages = "094507",
    year = "2019"
}

@article{Hanada:2023krw,
    author = "Hanada, Masanori and Ohata, Hiroki and Shimada, Hidehiko and Watanabe, Hiromasa",
    title = "{A New Perspective on Thermal Transition in QCD}",
    eprint = "2310.01940",
    archivePrefix = "arXiv",
    primaryClass = "hep-th",
    reportNumber = "YITP-23-121",
    doi = "10.1093/ptep/ptae044",
    journal = "PTEP",
    volume = "2024",
    number = "4",
    pages = "041B02",
    year = "2024"
}

@article{Fujimoto:2025sxx,
    author = "Fujimoto, Yuki and Fukushima, Kenji and Hidaka, Yoshimasa and McLerran, Larry",
    title = "{New state of matter between the hadronic phase and the quark-gluon plasma?}",
    eprint = "2506.00237",
    archivePrefix = "arXiv",
    primaryClass = "hep-ph",
    reportNumber = "N3AS-25-009, RIKEN-iTHEMS-Report-25, YITP-25-80",
    doi = "10.1103/h71y-km92",
    journal = "Phys. Rev. D",
    volume = "112",
    number = "7",
    pages = "074006",
    year = "2025"
}

@article{Mickley:2025qgk,
    author = "Mickley, Jackson A. and Allton, Chris and Bignell, Ryan and Leinweber, Derek B.",
    title = "{Novel insight into center-vortex geometry in four dimensions}",
    eprint = "2507.17200",
    archivePrefix = "arXiv",
    primaryClass = "hep-lat",
    reportNumber = "ADP25-26/T1288",
    doi = "10.1103/ptfd-yjtb",
    journal = "Phys. Rev. D",
    volume = "112",
    number = "5",
    pages = "054505",
    year = "2025"
}

@article{Wilson:1974sk,
    author = "Wilson, Kenneth G.",
    editor = "Taylor, J. C.",
    title = "{Confinement of Quarks}",
    reportNumber = "CLNS-262",
    doi = "10.1103/PhysRevD.10.2445",
    journal = "Phys. Rev. D",
    volume = "10",
    pages = "2445--2459",
    year = "1974"
}

@article{Mandelstam:1974pi,
    author = "Mandelstam, S.",
    title = "{Vortices and Quark Confinement in Nonabelian Gauge Theories}",
    reportNumber = "PRINT-74-1623 (UC,BERKELEY)",
    doi = "10.1016/0370-1573(76)90043-0",
    journal = "Phys. Rept.",
    volume = "23",
    pages = "245--249",
    year = "1976"
}

@inproceedings{tHooft:1975yol,
  author    = {Gerard 't Hooft},
  title     = {Gauge Theory for Strong Interactions},
  booktitle = {New Phenomena in Subnuclear Physics},
  editor    = {Antonio Zichichi},
  year      = {1976},
  pages     = {1225--1249},
  publisher = {Editrice Compositori},
  address   = {Bologna},
  note      = {Proceedings of the talk given at the 13th International School of Subnuclear Physics: New Phenomena in Subnuclear Physics, 11 July-1 August 1975. Erice, Italy}
}

@article{tHooft:1981bkw,
    author = "'t Hooft, Gerard",
    title = "{Topology of the Gauge Condition and New Confinement Phases in Nonabelian Gauge Theories}",
    reportNumber = "CALT-68-819",
    doi = "10.1016/0550-3213(81)90442-9",
    journal = "Nucl. Phys. B",
    volume = "190",
    pages = "455--478",
    year = "1981"
}

@article{Kronfeld:1987vd,
    author = "Kronfeld, Andreas S. and Schierholz, G. and Wiese, U. J.",
    title = "{Topology and Dynamics of the Confinement Mechanism}",
    reportNumber = "DESY-87-023",
    doi = "10.1016/0550-3213(87)90080-0",
    journal = "Nucl. Phys. B",
    volume = "293",
    pages = "461--478",
    year = "1987"
}

@article{Kronfeld:1987ri,
    author = "Kronfeld, Andreas S. and Laursen, M. L. and Schierholz, G. and Wiese, U. J.",
    title = "{Monopole Condensation and Color Confinement}",
    reportNumber = "DESY-87-073",
    doi = "10.1016/0370-2693(87)90910-5",
    journal = "Phys. Lett. B",
    volume = "198",
    pages = "516--520",
    year = "1987"
}

@article{Ivanenko:1991wt,
    author = "Ivanenko, T. L. and Pochinsky, A. V. and Polikarpov, M. I.",
    title = "{Condensate of Abelian monopoles and confinement in lattice gauge theories}",
    reportNumber = "LNF-91-086-P",
    doi = "10.1016/0370-2693(93)90427-J",
    journal = "Phys. Lett. B",
    volume = "302",
    pages = "458--462",
    year = "1993"
}

@article{Ezawa:1982bf,
    author = "Ezawa, Z. F. and Iwazaki, A.",
    title = "{Abelian Dominance and Quark Confinement in Yang-Mills Theories}",
    reportNumber = "TU-81-229-REV, TU-81-229",
    doi = "10.1103/PhysRevD.25.2681",
    journal = "Phys. Rev. D",
    volume = "25",
    pages = "2681",
    year = "1982"
}

@article{DelDebbio:1995yf,
    author = "Del Debbio, L. and Di Giacomo, A. and Paffuti, G. and Pieri, P.",
    title = "{Color confinement as dual Meissner effect: SU(2) gauge theory}",
    eprint = "hep-lat/9505014",
    archivePrefix = "arXiv",
    reportNumber = "IFUP-TH-22-95, SWAT-94-95-76",
    doi = "10.1016/0370-2693(95)00702-M",
    journal = "Phys. Lett. B",
    volume = "355",
    pages = "255--259",
    year = "1995"
}

@article{DiGiacomo:1999yas,
    author = "Di Giacomo, A. and Lucini, B. and Montesi, L. and Paffuti, G.",
    title = "{Color confinement and dual superconductivity of the vacuum. 1.}",
    eprint = "hep-lat/9906024",
    archivePrefix = "arXiv",
    reportNumber = "IFUP-TH-35-99",
    doi = "10.1103/PhysRevD.61.034503",
    journal = "Phys. Rev. D",
    volume = "61",
    pages = "034503",
    year = "2000"
}

@article{DiGiacomo:1999fb,
    author = "Di Giacomo, A. and Lucini, B. and Montesi, L. and Paffuti, G.",
    title = "{Color confinement and dual superconductivity of the vacuum. 2.}",
    eprint = "hep-lat/9906025",
    archivePrefix = "arXiv",
    reportNumber = "IFUP-TH-36-99",
    doi = "10.1103/PhysRevD.61.034504",
    journal = "Phys. Rev. D",
    volume = "61",
    pages = "034504",
    year = "2000"
}

@article{Chernodub:1996ps,
    author = "Chernodub, M. N. and Polikarpov, M. I. and Veselov, A. I.",
    title = "{Effective constraint potential for Abelian monopole in SU(2) lattice gauge theory}",
    eprint = "hep-lat/9610007",
    archivePrefix = "arXiv",
    reportNumber = "KANAZAWA-96-18",
    doi = "10.1016/S0370-2693(97)00309-2",
    journal = "Phys. Lett. B",
    volume = "399",
    pages = "267--273",
    year = "1997"
}

@article{Carmona:2001ja,
    author = "Carmona, J. M. and D'Elia, Massimo and Di Giacomo, A. and Lucini, B. and Paffuti, G.",
    title = "{Color confinement and dual superconductivity of the vacuum. 3.}",
    eprint = "hep-lat/0103005",
    archivePrefix = "arXiv",
    reportNumber = "IFUP-TH-2001-10",
    doi = "10.1103/PhysRevD.64.114507",
    journal = "Phys. Rev. D",
    volume = "64",
    pages = "114507",
    year = "2001"
}

@article{Chernodub:2006gu,
    author = "Chernodub, M. N. and Zakharov, V. I.",
    title = "{Magnetic component of Yang-Mills plasma}",
    eprint = "hep-ph/0611228",
    archivePrefix = "arXiv",
    reportNumber = "ITEP-LAT-2006-10, KANAZAWA-2006-16",
    doi = "10.1103/PhysRevLett.98.082002",
    journal = "Phys. Rev. Lett.",
    volume = "98",
    pages = "082002",
    year = "2007"
}

@article{DAlessandro:2007lae,
    author = "D'Alessandro, Alessio and D'Elia, Massimo",
    title = "{Magnetic monopoles in the high temperature phase of Yang-Mills theories}",
    eprint = "0711.1266",
    archivePrefix = "arXiv",
    primaryClass = "hep-lat",
    reportNumber = "GEF-TH-22-07",
    doi = "10.1016/j.nuclphysb.2008.03.002",
    journal = "Nucl. Phys. B",
    volume = "799",
    pages = "241--254",
    year = "2008"
}

@article{Cristoforetti:2009tx,
    author = "Cristoforetti, Marco and Shuryak, Edward",
    title = "{Bose-Einstein Condensation of strongly interacting bosons: From liquid He-4 to QCD monopoles}",
    eprint = "0906.2019",
    archivePrefix = "arXiv",
    primaryClass = "hep-ph",
    doi = "10.1103/PhysRevD.80.054013",
    journal = "Phys. Rev. D",
    volume = "80",
    pages = "054013",
    year = "2009"
}

@article{DAlessandro:2010jdd,
    author = "D'Alessandro, Alessio and D'Elia, Massimo and Shuryak, Edward V.",
    title = "{Thermal Monopole Condensation and Confinement in finite temperature Yang-Mills Theories}",
    eprint = "1002.4161",
    archivePrefix = "arXiv",
    primaryClass = "hep-lat",
    doi = "10.1103/PhysRevD.81.094501",
    journal = "Phys. Rev. D",
    volume = "81",
    pages = "094501",
    year = "2010"
}

@article{Greensite:2008ss,
    author = "Greensite, J. and Lucini, B.",
    title = "{Is Confinement a Phase of Broken Dual Gauge Symmetry?}",
    eprint = "0806.2117",
    archivePrefix = "arXiv",
    primaryClass = "hep-lat",
    doi = "10.1103/PhysRevD.78.085004",
    journal = "Phys. Rev. D",
    volume = "78",
    pages = "085004",
    year = "2008"
}

@article{Bonati:2011jv,
    author = "Bonati, Claudio and Cossu, Guido and D'Elia, Massimo and Di Giacomo, Adriano",
    title = "{The disorder parameter of dual superconductivity in QCD revisited}",
    eprint = "1111.1541",
    archivePrefix = "arXiv",
    primaryClass = "hep-lat",
    doi = "10.1103/PhysRevD.85.065001",
    journal = "Phys. Rev. D",
    volume = "85",
    pages = "065001",
    year = "2012"
}

@article{DeGrand:1980eq,
    author = "DeGrand, Thomas A. and Toussaint, Doug",
    editor = "Julve, J. and Ram{\'o}n-Medrano, M.",
    title = "{Topological Excitations and Monte Carlo Simulation of Abelian Gauge Theory}",
    reportNumber = "UCSB-TH-22-1980, NSF-ITP-80-36",
    doi = "10.1103/PhysRevD.22.2478",
    journal = "Phys. Rev. D",
    volume = "22",
    pages = "2478",
    year = "1980"
}

@article{Crean:2024nro,
    author = "Crean, Xavier and Giansiracusa, Jeffrey and Lucini, Biagio",
    title = "{Topological data analysis of monopole current networks in $U(1)$ lattice gauge theory}",
    eprint = "2403.07739",
    archivePrefix = "arXiv",
    primaryClass = "hep-lat",
    doi = "10.21468/SciPostPhys.17.4.100",
    journal = "SciPost Phys.",
    volume = "17",
    number = "4",
    pages = "100",
    year = "2024"
}

@article{Crean:2025gne,
    author = "Crean, Xavier and Giansiracusa, Jeffrey and Lucini, Biagio",
    title = "{Topological Data Analysis of Abelian Magnetic Monopoles in Gauge Theories}",
    eprint = "2501.19320",
    archivePrefix = "arXiv",
    primaryClass = "hep-lat",
    doi = "10.22323/1.466.0395",
    journal = "PoS",
    volume = "LATTICE2024",
    pages = "395",
    year = "2025"
}

@article{Bonati:2013bga,
    author = "Bonati, Claudio and D'Elia, Massimo",
    title = "{The Maximal Abelian Gauge in SU(N) gauge theories and thermal monopoles for N = 3}",
    eprint = "1308.0302",
    archivePrefix = "arXiv",
    primaryClass = "hep-lat",
    reportNumber = "IFUP-TH-2013-16",
    doi = "10.1016/j.nuclphysb.2013.10.004",
    journal = "Nucl. Phys. B",
    volume = "877",
    pages = "233--259",
    year = "2013"
}

@article{Bonati:2010tz,
    author = "Bonati, Claudio and Di Giacomo, Adriano and Lepori, Luca and Pucci, Fabrizio",
    title = "{Monopoles, abelian projection and gauge invariance}",
    eprint = "1002.3874",
    archivePrefix = "arXiv",
    primaryClass = "hep-lat",
    reportNumber = "IFUP-TH-2010-08, SISSA-13-2010-EP",
    doi = "10.1103/PhysRevD.81.085022",
    journal = "Phys. Rev. D",
    volume = "81",
    pages = "085022",
    year = "2010"
}

@article{Lucini:2003zr,
    author = "Lucini, Biagio and Teper, Michael and Wenger, Urs",
    title = "{The High temperature phase transition in SU(N) gauge theories}",
    eprint = "hep-lat/0307017",
    archivePrefix = "arXiv",
    reportNumber = "OUTP-03-19P",
    doi = "10.1088/1126-6708/2004/01/061",
    journal = "JHEP",
    volume = "2004",
    number = "01",
    pages = "061",
    year = "2004"
}

@article{Luscher:1993dy,
    author = "Luscher, Martin",
    title = "{A Portable high quality random number generator for lattice field theory simulations}",
    eprint = "hep-lat/9309020",
    archivePrefix = "arXiv",
    reportNumber = "DESY-93-133",
    doi = "10.1016/0010-4655(94)90232-1",
    journal = "Comput. Phys. Commun.",
    volume = "79",
    pages = "100--110",
    year = "1994"
}

@article{Ferrenberg:1989ui,
    author = "Ferrenberg, Alan M. and Swendsen, Robert H.",
    title = "{Optimized Monte Carlo analysis}",
    doi = "10.1103/PhysRevLett.63.1195",
    journal = "Phys. Rev. Lett.",
    volume = "63",
    pages = "1195--1198",
    year = "1989"
}

@article{Jay:2020jkz,
    author = "Jay, William I. and Neil, Ethan T.",
    title = "{Bayesian model averaging for analysis of lattice field theory results}",
    eprint = "2008.01069",
    archivePrefix = "arXiv",
    primaryClass = "stat.ME",
    reportNumber = "FERMILAB-PUB-20-374-T",
    doi = "10.1103/PhysRevD.103.114502",
    journal = "Phys. Rev. D",
    volume = "103",
    pages = "114502",
    year = "2021"
}

@article{Cabibbo:1982zn,
    author = "Cabibbo, N. and Marinari, E.",
    title = "{A New Method for Updating SU(N) Matrices in Computer Simulations of Gauge Theories}",
    doi = "10.1016/0370-2693(82)90696-7",
    journal = "Phys. Lett. B",
    volume = "119",
    pages = "387--390",
    year = "1982"
}

@incollection{gudhi:CubicalComplex
, author =  "Pawel Dlotko"
, title =   "Cubical complex"
, publisher =  "{GUDHI Editorial Board}"
, booktitle =   "{GUDHI} User and Reference Manual"
, url = "http://gudhi.gforge.inria.fr/doc/latest/group__cubical__complex.html"
, year =        2015
}

@unpublished{Cardinali:2021mfh,
    author = "Cardinali, Marco and D'Elia, Massimo and Pasqui, Andrea",
    title = "{Thermal monopole condensation in QCD with physical quark masses}",
    eprint = "2107.02745",
    archivePrefix = "arXiv",
    note = "arXiv:2107.02745",
    primaryClass = "hep-lat",
    month = "7",
    year = "2021"
}

@article{topology-hypothesis-with-PH,
  title = {Persistent homology analysis of phase transitions},
  author = {Donato, Irene and Gori, Matteo and Pettini, Marco and Petri, Giovanni and De Nigris, Sarah and Franzosi, Roberto and Vaccarino, Francesco},
  journal = {Phys. Rev. E},
  volume = {93},
  issue = {5},
  pages = {052138},
  numpages = {10},
  year = {2016},
  month = {May},
  publisher = {American Physical Society},
  doi = {10.1103/PhysRevE.93.052138},
  eprint = "1601.03641",
  archivePrefix = "arXiv",
  url = {https://link.aps.org/doi/10.1103/PhysRevE.93.052138}
}

@article{Tran:2021wyg,
    author = "Tran, Quoc Hoan and Chen, Mark and Hasegawa, Yoshihiko",
    title = "{Topological Persistence Machine of Phase Transitions}",
    eprint = "2004.03169",
    archivePrefix = "arXiv",
    primaryClass = "cond-mat.stat-mech",
    doi = "10.1103/PhysRevE.103.052127",
    journal = "Phys. Rev. E",
    volume = "103",
    pages = "052127",
    year = "2021"
}

@article{Kashiwa:2021ctc,
    author = "Kashiwa, Kouji and Hirakida, Takehiro and Kouno, Hiroaki",
    title = "{Persistent Homology Analysis for Dense QCD Effective Model with Heavy Quarks}",
    eprint = "2103.12554",
    archivePrefix = "arXiv",
    primaryClass = "hep-lat",
    doi = "10.3390/sym14091783",
    journal = "Symmetry",
    volume = "14",
    number = "9",
    pages = "1783",
    year = "2022"
}

@article{Olsthoorn:2020xzs,
    author = "Olsthoorn, Bart and Hellsvik, Johan and Balatsky, Alexander V.",
    title = "{Finding hidden order in spin models with persistent homology}",
    eprint = "2009.05141",
    archivePrefix = "arXiv",
    primaryClass = "cond-mat.stat-mech",
    doi = "10.1103/physrevresearch.2.043308",
    journal = "Phys. Rev. Res.",
    volume = "2",
    number = "4",
    pages = "043308",
    year = "2020"
}

@article{Cole:2020hjx,
    author = "Cole, Alex and Loges, Gregory J. and Shiu, Gary",
    title = "{Quantitative and interpretable order parameters for phase transitions from persistent homology}",
    eprint = "2009.14231",
    archivePrefix = "arXiv",
    primaryClass = "cond-mat.stat-mech",
    doi = "10.1103/PhysRevB.104.104426",
    journal = "Phys. Rev. B",
    volume = "104",
    number = "10",
    pages = "104426",
    year = "2021"
}

@article{Sale:2021xsq,
    author = "Sale, Nicholas and Giansiracusa, Jeffrey and Lucini, Biagio",
    title = "{Quantitative analysis of phase transitions in two-dimensional XY models using persistent homology}",
    eprint = "2109.10960",
    archivePrefix = "arXiv",
    primaryClass = "cond-mat.stat-mech",
    doi = "10.1103/PhysRevE.105.024121",
    journal = "Phys. Rev. E",
    volume = "105",
    number = "2",
    pages = "024121",
    year = "2022"
}

@article{Sehayek:2022lxf,
    author = "Sehayek, Dan and Melko, Roger G.",
    title = "{Persistent homology of Z2 gauge theories}",
    eprint = "2201.09856",
    archivePrefix = "arXiv",
    primaryClass = "cond-mat.stat-mech",
    doi = "10.1103/PhysRevB.106.085111",
    journal = "Phys. Rev. B",
    volume = "106",
    number = "8",
    pages = "085111",
    year = "2022"
}

@article{Sale:2022qfn,
    author = "Sale, Nicholas and Lucini, Biagio and Giansiracusa, Jeffrey",
    title = "{Probing center vortices and deconfinement in SU(2) lattice gauge theory with persistent homology}",
    eprint = "2207.13392",
    archivePrefix = "arXiv",
    primaryClass = "hep-lat",
    doi = "10.1103/PhysRevD.107.034501",
    journal = "Phys. Rev. D",
    volume = "107",
    number = "3",
    pages = "034501",
    year = "2023"
}

@article{Spitz:2022tul,
    author = "Spitz, Daniel and Urban, Julian M. and Pawlowski, Jan M.",
    title = "{Confinement in non-Abelian lattice gauge theory via persistent homology}",
    eprint = "2208.03955",
    archivePrefix = "arXiv",
    primaryClass = "hep-lat",
    doi = "10.1103/PhysRevD.107.034506",
    journal = "Phys. Rev. D",
    volume = "107",
    number = "3",
    pages = "034506",
    year = "2023"
}

@article{Spitz:2024bqh,
    author = "Spitz, Daniel and Urban, Julian M. and Pawlowski, Jan M.",
    title = "{Topological data analysis of the deconfinement transition in SU(3) lattice gauge theory}",
    eprint = "2412.09112",
    archivePrefix = "arXiv",
    primaryClass = "hep-lat",
    reportNumber = "MIT-CTP/5801",
    doi = "10.1103/k2xs-4y67",
    journal = "Phys. Rev. D",
    volume = "111",
    number = "11",
    pages = "114519",
    year = "2025"
}

@article{Nguyen:2024ikq,
    author = {Nguyen, Mendel and Sulejmanpasic, Tin and {\"U}nsal, Mithat},
    title = "{Phases of Theories with ZN 1-Form Symmetry, and the Roles of Center Vortices and Magnetic Monopoles}",
    eprint = "2401.04800",
    archivePrefix = "arXiv",
    primaryClass = "hep-th",
    doi = "10.1103/PhysRevLett.134.141902",
    journal = "Phys. Rev. Lett.",
    volume = "134",
    number = "14",
    pages = "141902",
    year = "2025"
}

@article{Giansiracusa:2025wqn,
    author = "Giansiracusa, Jeffrey and Lanners, David and Sulejmanpasic, Tin",
    title = "{Emergent Photons and Confinement: A Numerical Study on ZN Lattice Gauge Theory}",
    doi = "10.1103/h8mn-t4fk",
    journal = "Phys. Rev. Lett.",
    volume = "135",
    number = "22",
    pages = "221901",
    year = "2025"
}

@misc{ZENODO1,
    author = "Crean, Xavier and Giansiracusa, Jeffrey and Lucini, Biagio",
    title = "{Simplicity of confinement in $SU(3)$ Yang-Mills theory --- Data and Analysis Code Release}",
    year = "2026",
    eprint = " ",   
    doi = "10.5281/zenodo.18395360",
    archivePrefix = " ",
}

@misc{ZENODO2,
    author = "Crean, Xavier and Giansiracusa, Jeffrey and Lucini, Biagio",
    title = "{Simplicity of confinement in $SU(3)$ Yang-Mills theory --- Monte Carlo Code Release}",
    year = "2026",
    eprint = " ",   
    doi = "10.5281/zenodo.18171195",
    archivePrefix = " ",
}
